\documentclass[3p,times]{elsarticle}
\usepackage{hyperref}
\usepackage{amsmath}
\usepackage{amssymb}
\usepackage{amsthm}
\usepackage{mathrsfs}

\usepackage{graphicx}
\usepackage{xcolor}
\newcommand{\mb}{\boldsymbol}  
\journal{arXiv}

\begin{document}

\begin{frontmatter}

\title{New insight into light propagation and light-matter interactions \\ with applications to experimental observations}
\author{Changbiao Wang}
\ead{changbiao\_ wang@yahoo.com}
\address{ShangGang Group, 70 Huntington Road, Apartment 11, New Haven, CT 06512, USA}

\begin{abstract}
The paper provides a new understanding of light propagation and light-matter interactions by examining the physical implications of group velocity, electromagnetic (EM) power flow, Poynting theorem, energy conservation law, and Fermat's principle.  A criterion is set up to identify the justification of the group velocity definition, and a modified definition is proposed to remove the flaws that the classical definition has.  It is reasonably argued that energy conservation law and Fermat's principle are physical postulates independent of Maxwell equations.  A ``superluminal power flow'' is constructed to show that Poynting theorem cannot uniquely define the EM power flow if the energy conservation law or Fermat's principle is not taken into account.   As an application, associated basic concepts in textbooks and experimental observations reported in recent research works are also reviewed, including: why the traditional formulation of Fermat's principle has a limited application; how the Fermat's principle is formulated for a plane wave; why the Fermat's principle is consistent with Maxwell EM theory; what the significant difference is between Poynting theorem and energy conservation law; why Poynting vector as EM power flow may break energy conservation law and Fermat's principle in an anisotropic medium; why the physical explanations for ``spatially structured'' photons in Giovannini-coworkers experiments are not consistent with the principle of relativity; why the traditionally-argued invariance of information velocity contradicts Maxwell equations; and why the superluminal light pulse propagation claimed in Wang-Kuzmich-Dogariu experiments voilates Einstein causality. 
\end{abstract}

\begin{keyword}
light-matter interactions \sep group velocity \sep energy conservation \sep Fermat's principle \sep principle of relativity
%\PACS 03.50.De \sep 03.30.+p \sep 41.20.Jb
\end{keyword}

\end{frontmatter}
%\newpage
%% main text

%****Section 1*****
\section{Introduction}
\label{s1}
The paper provides a new understanding of light propagation and light-matter interactions by examining the physical implications of group velocity, electromagnetic power (energy) flow, Poynting theorem, energy conservation law, and Fermat's principle.  Associated basic concepts and experimental observations in past publications are reviewed.

It is a widely-accepted concept in the community that, group velocity is the transport velocity of electromagnetic (EM) signal energy.  For example, in their textbook Landau and Lifshitz state that ``the physical velocity of propagation of the waves is called the group velocity'' \cite[p.237]{r1};  for resolution of the Abraham-Minkowski debate on light momentum in a medium, Kemp argues that ``the pulse propagates at the group velocity'' \cite{r2}; for introducing a fundamental assumption in descriptions of light-matter interactions, Bethune-Waddell and Chau ``interpret the electromagnetic field as a fluid with an equivalent mass density ... moving with a group velocity'' \cite{r3}; in a recent experimental research report, Giovannini and coworkers claim that ``single photons travel at the group velocity'' \cite{r4}.   

The classical definition of group velocity is given by $\mathbf{v}_{\mathrm{gr-c}}=\partial\omega/\partial\textbf{k}_{\mathrm{w}}$, defined in the normal vector direction of $\textbf{k}_{\mathrm{w}}$-surface in the wave-vector space \cite[p.237]{r1} \cite{r5}, where $\omega$ is the angular frequency and $\textbf{k}_{\mathrm{w}}$ is the wave vector. 

\begin{figure} % figure 1
\begin{minipage}{\columnwidth}
\centering
%\framebox[\columnwidth][c]{\raisebox{0pt}[20mm][20mm]{figure image}}
\includegraphics[trim=1.2in 5.7in 1in 1.0in, clip=true,scale=0.54]{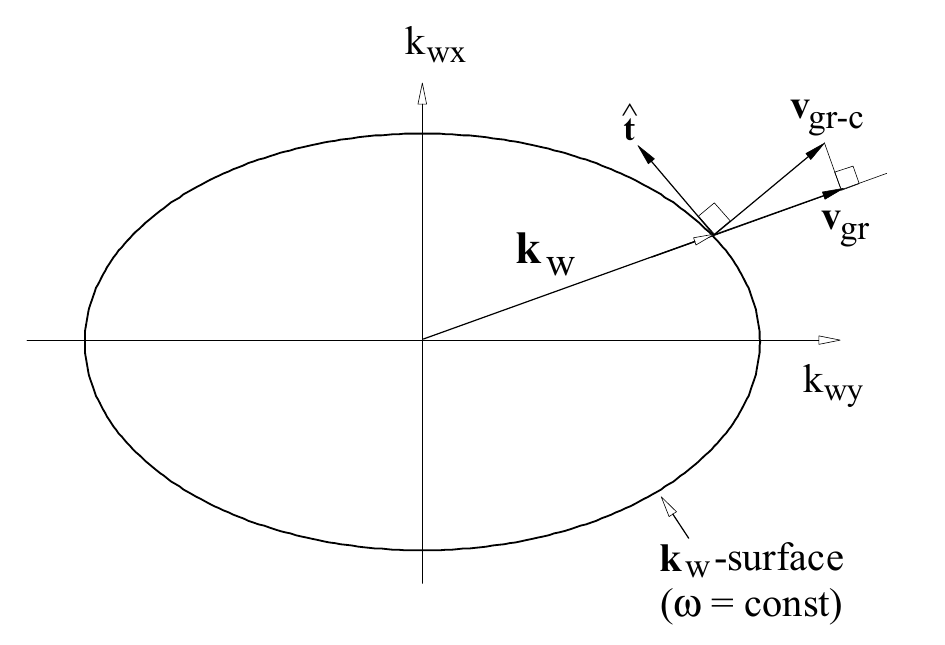}
\end{minipage}
\caption{Classical group velocity $\textbf{v}_{\mathrm{gr-c}}$ and modified group velocity $\textbf{v}_{\mathrm{gr}}$  in $\textbf{k}_{\mathrm{w}}$-space for an anisotropic medium.  $\textbf{v}_{\mathrm{gr-c}}$  is defined in the normal vector direction of the $\textbf{k}_{\mathrm{w}}$-surface; $\textbf{v}_{\mathrm{gr}}$ is defined in the  $\textbf{k}_{\mathrm{w}}$-direction. ~\textbf{\^t} is the unit tangential vector on the $\textbf{k}_{\mathrm{w}}$-surface.  For an isotropic medium, the $\textbf{k}_{\mathrm{w}}$-surface  becomes spherical, $\textbf{v}_{\mathrm{gr}}=\textbf{v}_{\mathrm{gr-c}}$  holds, and the two definitions are equivalent. }
\label{fig1}
\end{figure}

There are two flaws in the classical definition: (a) the group velocity can be greater than the phase velocity and break Fermat's principle for a plane wave in a \emph{non-dispersive, lossless, non-conducting}, anisotropic uniform medium; (b) the definition is not consistent with the principle of relativity for a plane wave in a moving isotropic uniform medium.  (The moving isotropic medium becomes anisotropic in general \cite{r5}.)  In this paper, we propose a modified group velocity definition, given by $\textbf{v}_{\mathrm{gr}}=\textbf{\^n}\partial\omega / \partial |\textbf{k}_{\mathrm{w}}| $ with $\textbf{\^n}=\textbf{k}_{\mathrm{w}}/|\textbf{k}_{\mathrm{w}}|$  the unit wave vector.  The modified group velocity is defined in the wave-vector direction, and it removes the above flaws.  The two definitions are shown in Fig.\,\ref{fig1}. 
%*****Figure 1*****
%*****Figure 1*****
Between the two definitions, $\textbf{v}_{\mathrm{gr}}\cdot\textbf{\^n}=\textbf{v}_{\mathrm{gr-c}}\cdot\textbf{\^n}$  holds.  For an isotropic medium, the $\textbf{k}_{\mathrm{w}}$-surface is spherical and the two definitions are equivalent. 

Strictly speaking, it is difficult to precisely define the velocity of EM energy transport for a \emph{practical} light pulse because of the existence of inevitable dispersions and divergences, more or less resulting in the pulse shape deformed.  However if a light pulse moves in a ``rigid'' way, then the EM energy velocity is exactly, without any ambiguity, equal to the rigidly-moving velocity of the pulse.  It is shown in the paper that, in a non-dispersive, lossless, non-conducting uniform medium (no matter whether it is isotropic or anisotropic), a plane-wave light pulse propagates rigidly at the modified group velocity.  Thus the modified group velocity can be taken to be an approximate description of EM energy velocity for practical situations with weak dispersions and low losses. 

Poynting theorem, called \emph{energy conservation equation}, is a specific mathematical description of energy conservation in electrodynamics, and it is an equation of \emph{differential form}; however, it is the differential form that results in the ambiguity of the definition of EM power (energy) flow. Fortunately, \emph{energy conservation law} is an \emph{independent} postulate in physics, while the principle of Fermat,\footnote
{\label{f1} Fermat's principle states that \emph{Nature always acts by the shortest course} \cite[p.xxi]{r6}.  When applied to optics, this principle requires light to take the path of least time.  This kind of description of Fermat's principle in optics is very general; however, how to use is really tricky.  Usually in textbooks \cite[p.291]{r1}\,\cite[p.363]{r5}\,\cite[p.128]{r6}, a specific formulation of Fermat's principle within the Maxwell-equation frame is about the optical path between two points, $A$ and $B$: How can a ray of light, emitted from point $A$, reach point $B$?  Mathematically, this form of Fermat's principle corresponds to a formulation of a variational principle ``which is weaker but which has a wider range of validity'' \cite[p.128]{r6}, and it has been generalized for light rays in general relativity \cite{r7,r8,r9} and nonstationary media \cite{r10}.  However, it should be indicated that in principle, this form of Fermat's principle is only applicable for a point light source at point $A$, and it is not applicable for a uniform plane wave used in the present paper.  To better understand this, let us take a simple example.  Suppose that there is a plane wave in free space, with the line $AB$ not parallel to the wave vector.  In such a case, the actual light ray never goes from $A$ to $B$ because the actual light ray must be perpendicular to the equiphase planes \cite[p.114]{r6}.  From this we can see the importance of understanding the exact implication of the general description of Fermat's principle: \emph{Light takes the path of least time}.  See footnote \ref{f4} for formulation of Fermat's principle for a plane wave.
} 
as a basic postulate of geometric optics, is not included in but consistent with Maxwell EM theory, and it is an additional physical condition that is imposed on the direction of EM energy transport.  Thus the ambiguity appearing in Poynting theorem can be clarified by energy conservation law or Fermat's principle.  In this paper, based on the Fermat's principle a criterion is set up to identify the justification of group velocity definition.  A ``superluminal power flow'' is constructed to show that the EM power flow cannot be uniquely defined by Poynting theorem without energy conservation law or Fermat's principle taken into account.  

As we know, formulation of any physical theories is usually restricted by fundamental physical postulates.  Thus a correct EM power flow is supposed to meet (i) EM energy conservation law, (ii) Fermat's principle, and (iii) the principle of relativity.  However traditionally, only (i) is emphasized through Poynting theorem, while (ii) and (iii) are neglected.  In this paper with all the above three postulates taken into account, a real EM power flow is presented. 
 
The paper is organized as follows.  In Sec.\,\ref{s2}, the classical group velocity is shown to break Fermat's principle and energy conservation law in a stationary symmetric anisotropic uniform medium.  In Sec.\,\ref{s3}, a criterion is set up to identify the justification of group velocity definition.  In Sec.\,\ref{s4}, the classical group velocity is shown to violate relativity principle in a moving isotropic uniform medium, and in Sec.\,\ref{s5}, some conclusions and remarks are given.  In \ref{appa}, it is shown that the modified group velocity is exactly equal to the rigidly-moving velocity of a plane-wave light pulse. In footnote \ref{f1}, it is shown why the traditional formulation of Fermat's principle has a limited application; in footnote \ref{f4}, how to formulate Fermat's principle for a plane wave is presented, and it is shown that the Fermat's principle is consistent with Maxwell EM theory; in footnote \ref{f5}, a subtle, but apparently significant difference between Poynting theorem and energy conservation law is clearly shown; in footnote \ref{f6}, it is shown why the physical explanations for ``spatially structured'' photons in Giovannini-coworkers experiments are not consistent with the principle of relativity; in footnote \ref{f10}, it is shown why the traditionally-argued invariance of information velocity is not consistent with Maxwell equations; and in footnote \ref{f11}, it is shown why the superluminal light pulse propagation reported in Wang-Kuzmich-Dogariu experimental demonstrations breaks Einstein causality.

%****Section 2*****
\section{Classical group velocity greater than phase velocity and breaking Fermat's principle}
\label{s2}
In this section, we will show that the classical group velocity can be greater than the phase velocity and break Fermat's principle for a signal plane wave in a non-dispersive, lossless, non-conducting, anisotropic uniform medium.  

A signal plane wave is composed of component monochromatic plane waves with different frequencies, which propagate in the same direction. A single plane-wave light pulse and a periodical plane-wave light pulse are typical examples of signal plane waves.  According to classical electrodynamics and Fourier analysis, the frequency spectrum is continuous for the former, while it is discrete for the latter.

 For a monochromatic plane wave with a phase function $\Psi=(\omega t-\textbf{k}_{\mathrm{w}}\cdot\textbf{x})$, Maxwell equations are simplified into  
\begin{align}
\omega\textbf{B}=\textbf{k}_{\mathrm{w}}\times\textbf{E}, \hspace{6mm} \omega\textbf{D}=-\textbf{k}_{\mathrm{w}}\times\textbf{H}, 
\label{eq1} 
\\
\textbf{k}_{\mathrm{w}}\cdot\textbf{B}=0,  \hspace{11mm}  \textbf{k}_{\mathrm{w}}\cdot\textbf{D}=0,~~~
\label{eq2}
\end{align} 
where $(\textbf{E}, \textbf{D}, \textbf{B}, \textbf{H}) = (\textbf{E}_0,\textbf{D}_0,\textbf{B}_0,\textbf{H}_0)\cos\Psi$ with $\textbf{E}_0$, $\textbf{D}_0$, $\textbf{B}_0$, and $\textbf{H}_0$  the real constant vectors; $\omega$  and $\textbf{k}_{\mathrm{w}}$  are real because the medium is assumed to be non-conducting and lossless.  

By making cross products of $\textbf{k}_{\mathrm{w}}\times(\omega\textbf{B}=\textbf{k}_{\mathrm{w}}\times\textbf{E})$ and $\textbf{k}_{\mathrm{w}}\times(\omega\textbf{D}=-\textbf{k}_{\mathrm{w}}\times\textbf{H})$ from Eq.\,(\ref{eq1}), with vector identity 
$\textbf{a}\times(\textbf{b}\times\textbf{c})=(\textbf{a}\cdot\textbf{c})\textbf{b}-(\textbf{a}\cdot\textbf{b})\textbf{c} $ taken into account, we have 
\begin{align}
\textbf{E}&=(\textbf{\^n}\cdot\textbf{E})\textbf{\^n}-\textbf{v}_{\mathrm{ph}}\times\textbf{B},
\label{eq3}
\\
\textbf{H}&=(\textbf{\^n}\cdot\textbf{H})\textbf{\^n}+\textbf{v}_{\mathrm{ph}}\times\textbf{D},
\label{eq4}
\end{align} 
where $\textbf{v}_{\mathrm{ph}}=\textbf{\^n}(\omega/|\textbf{k}_{\mathrm{w}}|)$ is the phase velocity.  The medium refractive index is defined as $n_{\mathrm{d}}=|\textbf{k}_{\mathrm{w}}|/|\omega/c|$ \cite{r1}, with $c$ the vacuum light speed, and thus the phase velocity also can be written as $\textbf{v}_{\mathrm{ph}}=\textbf{\^n}(\omega/|\omega|)(c/n_{\mathrm{d}})$.

By making inner products of $\textbf{H}\cdot(\omega\textbf{B}=\textbf{k}_{\mathrm{w}}\times\textbf{E})$  and $\textbf{E}\cdot(\omega\textbf{D}=-\textbf{k}_{\mathrm{w}}\times\textbf{H})$ from Eq.\,(\ref{eq1}), with $\textbf{H}\cdot(\textbf{k}_{\mathrm{w}}\times\textbf{E}) = \textbf{E}\cdot(-\textbf{k}_{\mathrm{w}}\times\textbf{H})$ taken into account we obtain $\textbf{E}\cdot\textbf{D}=\textbf{B}\cdot\textbf{H}$. Setting $\textbf{S}=\textbf{E}\times\textbf{H}$ (Poynting vector) and $W_{\mathrm{em}}=0.5(\textbf{E}\cdot\textbf{D}+\textbf{B}\cdot\textbf{H})$ (EM energy density), from Eqs.\,(\ref{eq3}) and (\ref{eq4}) we obtain \cite{r11}  
\begin{align}
\textbf{S}=\textbf{S}_{\mathrm{power}}+\textbf{S}_{\mathrm{pseu}},
\label{eq5}
\end{align}
where
\begin{align}
\textbf{S}_{\mathrm{power}}&=~\textbf{v}_{\mathrm{ph}}^2(\textbf{D}\times\textbf{B})=W_{\mathrm{em}}\textbf{v}_{\mathrm{ph}},
\label{eq6}
\\
\textbf{S}_{\mathrm{pseu}}~&=-(\textbf{v}_{\mathrm{ph}}\cdot\textbf{H})\textbf{B}-(\textbf{v}_{\mathrm{ph}}\cdot\textbf{E})\textbf{D}.
\label{eq7}
\end{align} 

From Eq.\,(\ref{eq2}), we know that both $\textbf{B}$ and $\textbf{D}$ are perpendicular to $\textbf{v}_{\mathrm{ph}}\parallel\textbf{k}_{\mathrm{w}}$, which is well discussed in the textbook by Kong \cite{r5}.  Thus $\textbf{S}_{\mathrm{power}}$ and $\textbf{S}_{\mathrm{pseu}}$ are perpendicular each other, leading to $|\textbf{S}|>|\textbf{S}_{\mathrm{power}}|$ for $\textbf{S}_{\mathrm{pseu}}\neq 0$, and we have 
\begin{equation}
\left|\frac{\textbf{S}}{W_{\mathrm{em}}}\right|>|\textbf{v}_{\mathrm{ph}}|.
\label{eq8}
\end{equation} 

Again from Eq.\,(\ref{eq1}), following Landau-Lifshitz approach \cite{r1} we have  
\begin{align}
(\delta\omega)\textbf{B}\cdot\textbf{H}=\delta\textbf{k}_{\mathrm{w}}\cdot(\textbf{E}\times\textbf{H})
-\omega(\delta\textbf{B})\cdot\textbf{H}+\omega(\textbf{D}\cdot\delta\textbf{E}),
\label{eq9}
\\
(\delta\omega)\textbf{D}\cdot\textbf{E}=\delta\textbf{k}_{\mathrm{w}}\cdot(\textbf{E}\times\textbf{H})
-\omega(\delta\textbf{D})\cdot\textbf{E}+\omega(\textbf{B}\cdot\delta\textbf{H}),
\label{eq10}
\end{align} 
where $\delta\textbf{k}_{\mathrm{w}}$  is an arbitrary infinitesimal change in wave vector.  From above Eq. (\ref{eq9}) and (\ref{eq10}), we obtain 
\begin{equation}
\delta\omega=\delta\textbf{k}_{\mathrm{w}}\cdot\frac{\textbf{S}}{W_{\mathrm{em}}}-\frac{\omega\Delta}{2W_{\mathrm{em}}},
\label{eq11}
\end{equation} 
where 
\begin{equation}
\Delta=(\delta\textbf{D}\cdot\textbf{E}-\textbf{D}\cdot\delta\textbf{E})+
(\delta\textbf{B}\cdot\textbf{H}-\textbf{B}\cdot\delta\textbf{H}).
\label{eq12}
\end{equation} \\
\indent Supposing that the dielectric constant tensors for the non-dispersive medium are symmetric (symmetric medium), we have $\delta\textbf{D}\cdot\textbf{E}-\textbf{D}\cdot\delta\textbf{E}=0$ and $\delta\textbf{B}\cdot\textbf{H}-\textbf{B}\cdot\delta\textbf{H}=0$ holding \cite{r1}. Inserting them into Eq.\,(\ref{eq12}), we have $\Delta=0$ holding for any $\delta\textbf{k}_{\mathrm{w}}$.  Then inserting $\Delta=0$ into Eq.\,(\ref{eq11}), we have $\delta\omega=\delta\textbf{k}_{\mathrm{w}}\cdot(\textbf{S}/{W_{\mathrm{em}}})$ holding for any $\delta\textbf{k}_{\mathrm{w}}$.  With the help of standard calculus,\footnote
{\label{f2}  If $u=u(x,y,z)$ is a differentiable function and $\mathrm{d}u=\textbf{A}(x,y,z)\cdot \mathrm{d}\textbf{x}$ holds for any $\textbf{x}$ and $\mathrm{d}\textbf{x}$, then we have $\nabla u=\textbf{A}$.  That is because $\mathrm{d}u=\nabla u\cdot \mathrm{d}\textbf{x}$ holds for any $\textbf{x}$ and $\mathrm{d}\textbf{x}$, and so does $(\textbf{A}-\nabla u)\cdot \mathrm{d}\textbf{x}=0$, namely $(\textbf{A}-\nabla u)\perp \mathrm{d}\textbf{x}$ for any $\mathrm{d}\textbf{x}$.  Thus $\textbf{A}-\nabla u\equiv 0$ holds for any $\textbf{x}$.
} 
we obtain the classical group velocity holding for any instantaneous time (instead of time average\footnote
{\label{f3}  In Chapter 11 of the book Ref.\,\cite{r12}, a more general result was shown for a dispersive lossless uniform medium based on a ``time average'' approach.  Usually the holding of time average $<\textbf{v}_{\mathrm{gr-c}}>=<\textbf{S}/W_{\mathrm{em}}>$ does not necessarily means the instant $\textbf{v}_{\mathrm{gr-c}}=\textbf{S}/W_{\mathrm{em}}$ holding; see Problem 1 on p. 353 of Ref.\,\cite{r1}.
}) \cite{r12}, given by 
\begin{equation}
\textbf{v}_{\mathrm{gr-c}}\equiv\frac{\partial\omega}{\partial\textbf{k}_{\mathrm{w}}}=\frac{\textbf{S}}{W_{\mathrm{em}}} \hspace{4mm}
\left( \parbox{30mm}{for a non-dispersive  symmetric medium} \right),
\label{eq13}
\end{equation} \\
where $\textbf{S}/W_{\mathrm{em}}$ is defined as the velocity of energy transport traditionally \cite[p.669]{r6}\cite{r13}, termed ``classical energy velocity'' for the convenience.  

Comparing Eq.\,(\ref{eq13}) with Eq.\,(\ref{eq8}), we find that the classical group velocity $|\textbf{v}_{\mathrm{gr-c}}|~(=|\textrm{classical energy velocity}|)$ is greater than the phase velocity $|\textbf{v}_{\mathrm{ph}}|$ for waves with $\textbf{S}_{\mathrm{pseu}} \neq 0$ in a non-dispersive anisotropic symmetric medium.  However according to Fermat's principle, the energy velocity for a monochromatic plane wave is equal to the phase velocity $\textbf{v}_{\mathrm{ph}}$.\footnote
{\label{f4}  \emph{Formulation of Fermat's principle for a plane wave.}  The principle of Fermat is an additional physical condition imposed on the direction of energy transport.  As we have known from footnote 1, a light ray proceeding from point $A$ in general does not necessarily pass through point $B$; however, a light ray proceeding from one equiphase surface must intersect another equiphase surface \cite[p.115]{r6}.  Thus the Fermat's principle for a plane wave can be formulated as: \emph{from one equiphase plane to the next}, the optical length of an actual ray is the shortest; namely, the actual ray is the one to make the optical length $\int n_{\mathrm{d}}\mathrm{d}s$ the minimum. The medium, which supports a plane wave, is uniform ($\partial n_{\mathrm{d}}/\partial\textbf{x}=0$), but it can be isotropic or anisotropic.  When $\int \mathrm{d}s$  is equal to the distance between the equiphase planes, $\int n_{\mathrm{d}}\mathrm{d}s=n_{\mathrm{d}}\int \mathrm{d}s$  reaches the minimum.  Thus the actual ray or the direction of energy transport must be parallel to the wave vector.  On the other hand, from Eq.\,(\ref{eq5}) we know that $\textbf{S}_{\mathrm{power}}$ is parallel to the wave vector while $\textbf{S}_{\mathrm{pseu}}$ is perpendicular to the wave vector.  If $\textbf{S}_{\mathrm{power}}=W_{\mathrm{em}}\textbf{v}_{\mathrm{ph}}$ is defined as the power flow, Fermat's principle is automatically satisfied, resulting in the EM energy velocity $:=\textbf{S}_{\mathrm{power}}/W_{\mathrm{em}}=\textbf{v}_{\mathrm{ph}}$ (phase velocity).  In addition, $\textbf{S}_{\mathrm{power}}$ satisfies the energy conservation equation $\nabla\cdot\textbf{S}_{\mathrm{power}}+\partial W_{\mathrm{em}}/\partial t=0$.  Thus we conclude that the Fermat's principle, presented above, is completely consistent with Maxwell EM theory. In fact, within the Maxwell-equation frame without Fermat's principle taken into account but with the energy conservation law imposed, one should have been aware that $\textbf{S}_{\mathrm{pseu}}$ is not of power flow because of $\nabla\cdot\textbf{S}_{\mathrm{pseu}}\equiv 0$, never responsible for power flowing at any places for any time.  Furthermore, it is worthwhile to emphasize that the Fermat's principle for a plane wave, formulated in the present paper, is applicable in both isotropic and anisotropic dielectric media. According to this formulation of Fermat's principle, the actual ray of light is in the wave-vector direction (instead of the Poynting-vector direction), no matter whether in an isotropic or anisotropic medium. In contrast, in classical textbooks, the light ray for a plane wave in an anisotropic medium is given in the Poynting-vector direction (instead of the wave-vector direction). For example, Landau and Lifshitz argue that ``the direction of the light rays (in geometrical optics) is given by the group velocity vector'' \cite[p.335]{r1} and ``the group velocity is in the same direction as the Poynting vector'' \cite[p.336]{r1}. Born and Wolf also argue that ``the ray velocity, is in the same direction as the Poynting vector'', and the ray velocity is equal to the (classical) energy velocity $\textbf{S}/W_{\mathrm{em}}$ \cite[p.669]{r6}. Straightforwardly speaking, in the textbooks \cite{r1,r6}, the light ray for a plane wave in an anisotropic medium is \emph{not} defined according to Fermat's principle; instead, it is defined based on a thought-to-be well-established but actually disproved basic concept that the Poynting vector always represents a real power (energy) flow, even in an anisotropic medium \cite{r11,r14}.
}  
Thus the result of Eq.\,(\ref{eq13}), $|\textbf{v}_{\mathrm{gr-c}}|=|\textrm{classical energy velocity}|>|\textbf{v}_{\mathrm{ph}}|$, is not consistent with the Fermat's principle, because all component monochromatic plane waves with different frequencies in a signal plane wave have the same phase velocity and thus the same energy velocity in a ``non-dispersive'' medium.  

However the modified definition does not have such an inconsistency, as shown as follows.  From the modified definition $\textbf{v}_{\mathrm{gr}}=$ $\textbf{\^n}\partial\omega/\partial|\textbf{k}_{\mathrm{w}}|$ with  $\textbf{v}_{\mathrm{ph}}=\textbf{\^n}(\omega/|\omega|)(c/n_{\mathrm{d}})$  taken into account, no matter whether the medium is symmetric or not, we have (see \ref{appa}) 
\begin{equation}
\textbf{v}_{\mathrm{gr}}=\frac{\textbf{v}_{\mathrm{ph}}}{1+(\omega/n_{\mathrm{d}})(\partial n_{\mathrm{d}}/\partial\omega)} \hspace{3mm}
\left( \parbox{23mm}{for a dispersive medium} \right).
\label{eq14}
\end{equation} \\
Equation (\ref{eq14}) has the same form as that for an isotropic medium \cite{r15}, and the only difference is that $n_{\mathrm{d}}$  is dependent on the direction of propagation of waves for the anisotropic medium, while it is not for an isotropic medium.

When no dispersion is involved ($\partial n_{\mathrm{d}}/\partial\omega=0$), we have $\textbf{v}_{\mathrm{gr}}=\textbf{v}_{\mathrm{ph}}$, namely the group velocity is equal to the phase velocity.  Thus the modified definition removes the above physical inconsistency in the classical definition.

From Eq.\,(\ref{eq6}) we have $\textbf{v}_{\mathrm{ph}}=\textbf{S}_{\mathrm{power}}/W_{\mathrm{em}}$ (= energy velocity).  Thus $\textbf{v}_{\mathrm{gr}}~(=\textbf{v}_{\mathrm{ph}})$ is also equal to the energy velocity $\textbf{S}_{\mathrm{power}}/W_{\mathrm{em}}$, completely in agreement with the criterion of group velocity definition (see Sec.\,\ref{s3}).

The problem ($|\textbf{v}_{\mathrm{gr-c}}|>|\textbf{v}_{\mathrm{ph}}|\textrm{ for }\textbf{S}_{\mathrm{pseu}}\neq 0$) in the classical definition of group velocity comes from the fact that in Eq.\,(\ref{eq5}), due to $\nabla\cdot\textbf{S}_{\mathrm{pseu}}\equiv 0$ \cite{r11,r14},  $\textbf{S}_{\mathrm{pseu}}$ is not responsible for any EM power flowing at any places for any time. Otherwise, energy conservation \emph{law} (instead of energy conservation \emph{equation}) would be broken, because ~$\textbf{S}_{\mathrm{power}}=W_{\mathrm{em}}\textbf{v}_{\mathrm{ph}}$ has already carried all the EM energy $W_{\mathrm{em}}$ propagating at $\textbf{v}_{\mathrm{ph}}$.  Thus we conclude that only $\textbf{S}_{\mathrm{power}}$ is a real power flow while $\textbf{S}_{\mathrm{pseu}}$  is a pseudo-power flow.\footnote
{\label{f5} \emph{Poynting theorem versus energy conservation law.}   ~From $\nabla\cdot\textbf{B}=0\Rightarrow\textbf{B}\cdot\textbf{k}_{\mathrm{w}}=0\Rightarrow\textbf{B}_0\cdot\textbf{k}_{\mathrm{w}}=0$    and $\nabla\cdot\textbf{D}=0\Rightarrow\textbf{D}\cdot\textbf{k}_{\mathrm{w}}=0\Rightarrow\textbf{D}_0\cdot\textbf{k}_{\mathrm{w}}=0$, we have $\nabla\cdot\textbf{S}_{\mathrm{pseu}}=2\cos\Psi\sin\Psi~[-(\textbf{v}_{\mathrm{ph}}\cdot\textbf{H}_0)\textbf{B}_0
-(\textbf{v}_{\mathrm{ph}}\cdot\textbf{E}_0)\textbf{D}_0]\cdot\textbf{k}_{\mathrm{w}}=0$ holding for any time and places, namely $\nabla\cdot\textbf{S}_{\mathrm{pseu}}\equiv 0$. Mathematically, $\textbf{S}_{\mathrm{pseu}}$ can carry a time-independent EM energy $W_{\mathrm{pseu}}$, namely $-\nabla\cdot\textbf{S}_{\mathrm{pseu}}=\partial W_{\mathrm{pseu}}/\partial t\equiv 0$. However $\textbf{S}_{\mathrm{power}}=W_{\mathrm{em}}\textbf{v}_{\mathrm{ph}}$, satisfying $-\nabla\cdot\textbf{S}_{\mathrm{power}}=\partial W_{\mathrm{em}}/\partial t$, already carries all the EM energy $W_{\mathrm{em}}$ propagating at $\textbf{v}_{\mathrm{ph}}$.  According to energy conservation law, we have $W_{\mathrm{em}}+W_{\mathrm{pseu}}=W_{\mathrm{em}}\Rightarrow W_{\mathrm{pseu}}=0$.  Thus in the frame of Maxwell EM theory with energy conservation law considered, we conclude that $\textbf{S}_{\mathrm{pseu}}$ does not transport any energy at any places for any time, and $\textbf{S}_{\mathrm{pseu}}$ is a pseudo-power flow.  Note that $W_{\mathrm{em}}+W_{\mathrm{pseu}}=W_{\mathrm{em}}\Rightarrow W_{\mathrm{pseu}}=0$ comes from energy conservation law, instead of energy conservation equation.  Poynting theorem or energy conservation equation only tells us that   $-\nabla\cdot\textbf{S}=\partial W_{\mathrm{em}}/\partial t \Rightarrow$ $-\nabla\cdot\textbf{S}_{\mathrm{power}}=\partial W_{\mathrm{em}}/\partial t$ and $-\nabla\cdot\textbf{S}_{\mathrm{pseu}}=\partial W_{\mathrm{pseu}}/\partial t\equiv 0$, but never tells us that $W_{\mathrm{em}}+W_{\mathrm{pseu}}=W_{\mathrm{em}}$ should be holding.  From here we can see why the energy conservation law is not included in but is consistent with Maxwell EM theory, and it is an independent postulate in physics.
} 
In other words, Poynting vector $\textbf{S}=\textbf{S}_{\mathrm{power}}+\textbf{S}_{\mathrm{pseu}}$ in Eq.\,(\ref{eq13}) does not represent the real power flow for waves with  $\textbf{S}_{\mathrm{pseu}} \neq 0$. 

It is interesting to point out that the above conclusion, obtained from the Maxwell EM theory with energy conservation law considered (as shown in footnote \ref{f5}), is completely consistent with the Fermat's principle formulated in footnote \ref{f4}.  That is to say,  both energy conservation law and Fermat's principle require that only $\textbf{S}_{\mathrm{power}}$ be the real EM power flow while $\textbf{S}_{\mathrm{pseu}}$ cannot be of EM power flow at all, because $\textbf{S}_{\mathrm{power}}$ already carries all EM energy (footnote \ref{f5}) propagating along the wave vector (footnote \ref{f4}). 

%****Section 3*****
\section{Criterion for identifying the justification of group velocity definition}
\label{s3}
As mentioned before, the group velocity is widely accepted as the velocity of signal energy transport in a medium.  Thus a rule or criterion is needed to identify the justification of definition.  Below we will show that there is such a criterion.

Suppose that the EM field is Einstein-light-quantized for a periodical signal plane wave that propagates in a non-dispersive, lossless, non-conducting, uniform medium ($\partial n_{\mathrm{d}}/ \partial \omega$ $=0$ and $\partial n_{\mathrm{d}}/\partial\textbf{x}=0$). All photons with different frequencies (energies) move in the same direction.  Physically, only groups of photons can perform signal transport.  The average velocity of a group of photons which constitutes a period of signals can be defined as the group velocity.  Due to no dispersion, all photons have the same velocity, and the energy of any group of such photons is transported at the photon's moving velocity.  Thus the group velocity is equal to the energy velocity in a non-dispersive medium for the signal plane wave.

On the other hand, the energy velocity of a signal plane wave also can be examined from the relation between photon's moving velocity and the phase velocity in the non-dispersive medium. 
 
First consider a monochromatic plane wave, of which the energy velocity is equal to the photon moving velocity because the photon is the carrier of energy.  The plane-wave phase function defines all equiphase planes of motion, with the wave vector as their normal vector.  From one equiphase plane to another equiphase plane, the path parallel to the normal vector is the shortest.  According to the Fermat's principle, the photon must move parallel to the wave vector, with its phase $\Psi=(\omega t-\textbf{k}_{\mathrm{w}}\cdot\textbf{x})$ kept unchanged.  Thus the photon velocity, namely the energy velocity of the monochromatic plane wave, is equal to the phase velocity.

A signal plane wave is made of component monochromatic waves with different frequencies but propagating in the same direction, and because of no dispersion, all these component waves have the same phase velocity, and further, they have the same energy velocity according to the above result for a monochromatic wave.  Thus the energy velocity of signal wave is equal to the phase velocity in a non-dispersive medium.  

From above analysis we conclude that the group, energy, and phase velocities are equal for a signal plane wave in a non-dispersive uniform medium no matter if it is isotropic or anisotropic, and this can be used as a criterion to identify the justification of definitions of group velocity. 

As we have known, derived from the classical group velocity definition $\textbf{v}_{\mathrm{gr-c}}=\partial\omega/\partial\textbf{k}_{\mathrm{w}}$, Eq.\,(\ref{eq13}) does not satisfy the above criterion.  In contrast, derived from the modified group velocity definition $\textbf{v}_{\mathrm{gr}}=\textbf{\^n}\partial\omega/\partial |\textbf{k}_{\mathrm{w}}|$, Eq.\,(\ref{eq14}) satisfies the criterion.  Thus the modified definition $\textbf{v}_{\mathrm{gr}}=\textbf{\^n}\partial\omega/\partial |\textbf{k}_{\mathrm{w}}|$ is justifiable.

%****Section 4*****
\section{Classical group velocity not consistent with the principle of relativity}
\label{s4}
We have shown that the group, energy, and phase velocities are equal for a signal plane wave in a non-dispersive, lossless, non-conducting, uniform medium no matter if it is isotropic or anisotropic.  According to the principle of relativity, this property is valid in all inertial frames.

The group velocity formula $\textbf{v}_{\mathrm{gr-c}}=\textbf{S}/W_{\mathrm{em}}$ given by Eq.\,(\ref{eq13}) was derived from the classical definition $\textbf{v}_{\mathrm{gr-c}}=\partial\omega/\partial\textbf{k}_{\mathrm{w}}$ and  $\Delta=0$ for a non-dispersive symmetric medium which is stationary.  Below we will show that, $\Delta=0$  and $\textbf{v}_{\mathrm{gr-c}}=\textbf{S}/W_{\mathrm{em}}$ also hold in all inertial frames for a \emph{moving} non-dispersive isotropic uniform medium, but $\textbf{v}_{\mathrm{gr-c}}=\textbf{v}_{\mathrm{ph}}$ doesn't; thus the classical definition $\textbf{v}_{\mathrm{gr-c}}=\partial\omega/\partial\textbf{k}_{\mathrm{w}}$ is not consistent with the principle of relativity.

\begin{figure} % figure 2
\begin{minipage}{\columnwidth}
\centering
\includegraphics[trim=1.2in 6.2in 1.3in 1.1in, clip=true,scale=0.5]{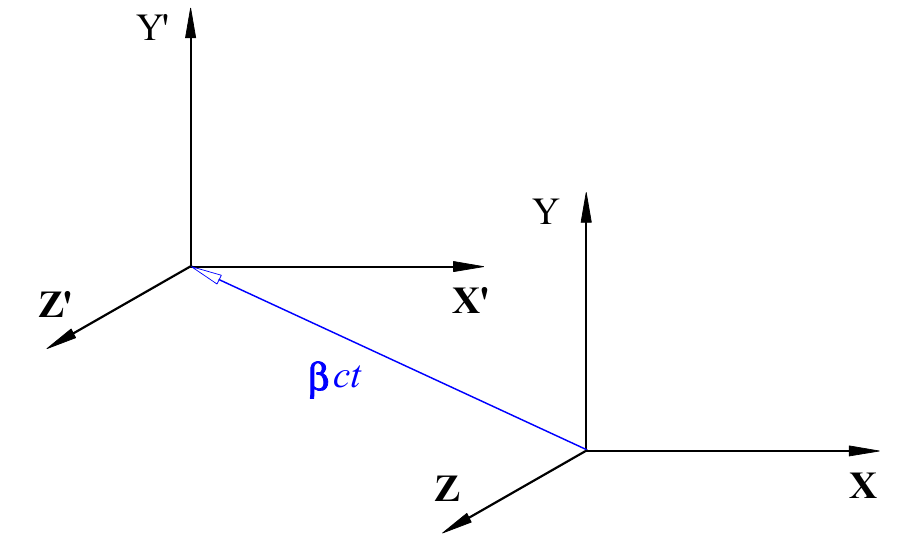}
\end{minipage}
\caption{Medium-rest frame $X'Y'Z'$ moves with respect to laboratory frame $XYZ$ at $\mb{\beta}c$, while $XYZ$ moves with respect to $X'Y'Z'$ at $\mb{\beta}'c$ (not shown), with ${\mb{\beta}'=-\mb{\beta}}$. Observed in the laboratory frame, the dielectric constant tensors are not symmetric, unless the medium moves along the wave vector, namely the moving isotropic uniform medium is an asymmetric anisotropic uniform medium. }
\label{fig2}
\end{figure}

Suppose that the medium is fixed in the frame $X'Y'Z'$, which moves at $\mb{\beta}c$ with respect to the laboratory frame $XYZ$, as shown in Fig.\,\ref{fig2}. 
%****Figure 2*****
%****Figure 2*****
From the wave four-vector $K^{\mu}=(\textbf{k}_{\mathrm{w}},\omega/c)$ we can obtain the Lorentz transformation of refractive index, given by \cite{r11} \\
\begin{equation}
n_{\mathrm{d}}=\frac{ \sqrt{ (n_{\mathrm{d}}'^2-1)+\gamma^2 (1-n_{\mathrm{d}}' \textbf{\^n}' \cdot\mb{\beta}')^2} }{|\gamma(1-n_{\mathrm{d}}'\textbf{\^n}' \cdot\mb{\beta}')|},
\label{eq15}
\end{equation} \\
where $\textbf{\^n}'$ is the unit wave vector and $n'_{\mathrm{d}}$ the refractive index in the medium-rest frame, and $\gamma=(1-\mb{\beta}^2)^{-1/2}$.  ~$n_{\mathrm{d}}$ is anisotropic while $n'_{\mathrm{d}}$ is isotropic.  Based on above Eq.\,(\ref{eq15}), first we will show below that if the medium has no dispersion observed in the medium-rest frame, then it does not have dispersion either observed in the laboratory frame.

Suppose that a signal plane wave propagates along the $\textbf{\^n}'$-direction in a uniform medium which has no dispersion ($\partial n'_{\mathrm{d}}/\partial\omega'=0$).  Observed in the medium-rest frame, all component waves with different frequencies propagate in the same direction, and they have the same refractive index $n'_{\mathrm{d}}$.  In terms of Eq.\,(\ref{eq15}), observed in the laboratory frame, all component waves with different frequencies also have the same refractive index $n_{\mathrm{d}}$.  Thus the medium has no dispersion either observed in the laboratory frame.  In other words, a moving non-dispersive medium is also non-dispersive ($\partial n_{\mathrm{d}}/\partial\omega=0$) observed in all inertial frames.  Thus from Eq.\,(\ref{eq14}) we have the modified group velocity  
\begin{equation}
\textbf{v}_{\mathrm{gr}}=\textbf{v}_{\mathrm{ph}} \hspace{3mm} \left( \parbox{42mm}{for a moving non-dispersive uniform medium} \right).
\label{eq16}
\end{equation}  \\ 
\indent From EM field Lorentz transformations for a plane wave, we have \cite{r11}   \vspace{2mm}
\begin{align}
\left[ \begin{array}{c} \textbf{E} \\ \textbf{H} \end{array} \right]&=\gamma \left(1-n'_{\mathrm{d}}\textbf{\^n}'\cdot\mb{\beta}'\right)\left[ \begin{array}{c} \textbf{E}' \\ \textbf{H}' \end{array} \right] +~\left( \gamma n'_{\mathrm{d}}\textbf{\^n}'-\frac{\gamma-1}{\mb{\beta}^2}\mb{\beta}'\right)
\left[ \begin{array}{c} \mb{\beta}'\cdot\textbf{E}' \\ \mb{\beta}'\cdot\textbf{H}' \end{array} \right],
\label{eq17}
\\
&~~
\notag \\
\left[ \begin{array}{c} \textbf{B} \\ \textbf{D} \end{array} \right]&=\gamma \left(1-\frac{1}{n'_{\mathrm{d}}}\textbf{\^n}'\cdot\mb{\beta}'\right)\left[ \begin{array}{c} \textbf{B}' \\ \textbf{D}' \end{array} \right] +~\left( \gamma \frac{1}{n'_{\mathrm{d}}}\textbf{\^n}'-\frac{\gamma-1}{\mb{\beta}^2}\mb{\beta}'\right)
\left[ \begin{array}{c} \mb{\beta}'\cdot\textbf{B}' \\ \mb{\beta}'\cdot\textbf{D}' \end{array} \right].
\label{eq18}
\end{align} \\
%****Figure 3*****
\begin{figure} % figure 3
\begin{minipage}{\columnwidth}
\centering
\includegraphics[trim=1.0in 4.7in 1in 1.1in, clip=true,scale=0.45]{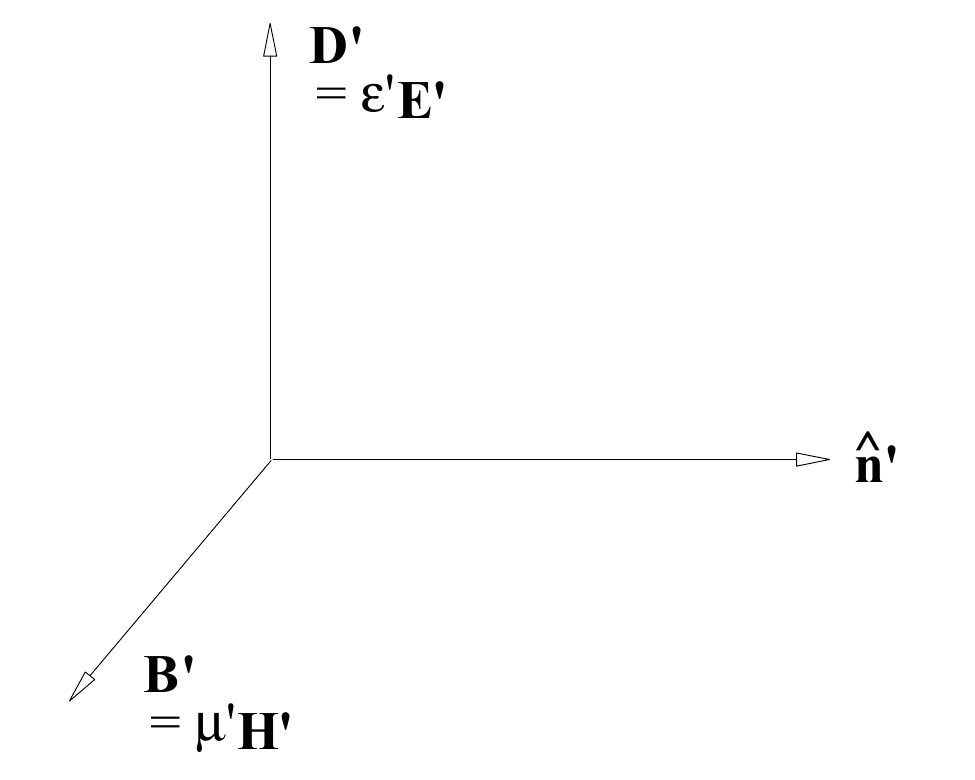}
\end{minipage}
\caption{EM fields $\textbf{E}'$, $\textbf{D}'$, $\textbf{B}'$, $\textbf{H}'$, and the unit wave vector  $\textbf{\^n}'$ in the medium-rest frame.  $\mathbf{D}'\perp\mathbf{B}'$ and $(\mathbf{D}'\times\mathbf{B}')\parallel\textbf{\^n}'$ hold.  Observed in the medium-rest frame, the uniform medium is isotropic, and $\textbf{D}'=\epsilon'\textbf{E}'$ and $\textbf{B}'=\mu'\textbf{H}'$ hold, with $\epsilon'$ and $\mu'$ being the scalar dielectric constants.   Since $\textbf{\^n}'\cdot\textbf{\^n}'=1 \Rightarrow \delta\textbf{\^n}'\cdot\textbf{\^n}'=0$ holds, $\delta\textbf{\^n}'$ is perpendicular to $\textbf{\^n}'$, and it is parallel to the $\textbf{D}'\text{--}\textbf{B}'$ plane.}
\label{fig3} 
\end{figure}
%****Figure 3*****
\indent Because the dielectric medium is isotropic observed in the medium-rest frame, and it is non-dispersive, $n'_{\mathrm{d}}$ is not dependent on the frequency and the direction of wave vector, leading to $\delta n'_{\mathrm{d}}=0$.  From Eqs.\,(\ref{eq17}) and (\ref{eq18}), we have 
\begin{align}
\delta\textbf{D}\cdot\textbf{E}-\textbf{D}\cdot\delta\textbf{E}&=\gamma^2 \left( -\frac{1}{n'_{\mathrm{d}}}+n'_{\mathrm{d}} \right) ~\delta\textbf{\^n}' \cdot \left[~\mb{\beta}'(\textbf{D}'\cdot\textbf{E}')-2\textbf{E}'(\mb{\beta}'\cdot\textbf{D}')\right],
\label{eq19}
\\
\delta\textbf{B}\cdot\textbf{H}-\textbf{B}\cdot\delta\textbf{H}&=\gamma^2 \left( -\frac{1}{n'_{\mathrm{d}}}+n'_{\mathrm{d}} \right) ~\delta\textbf{\^n}' \cdot \left[~\mb{\beta}'(\textbf{B}'\cdot\textbf{H}')-2\textbf{H}'(\mb{\beta}'\cdot\textbf{B}')\right].
\label{eq20}
\end{align}
Inserting Eq.\,(\ref{eq19}) and (\ref{eq20}) into Eq.\,(\ref{eq12}), with $\textbf{E}'\cdot\textbf{D}'=\textbf{B}'\cdot\textbf{H}'$ taken into account we have \\
\begin{equation}
\Delta=2\gamma^2 \left( -\frac{1}{n'_{\mathrm{d}}}+n'_{\mathrm{d}} \right)\delta\textbf{\^n}'\cdot\mb{\Pi}', 
\label{eq21}
\end{equation} 
where 
\begin{equation}
\mb{\Pi}'=\mb{\beta}'(\textbf{D}'\cdot\textbf{E}')-\textbf{E}'(\mb{\beta}'\cdot\textbf{D}')
-\textbf{H}'(\mb{\beta}'\cdot\textbf{B}').
\label{eq22}
\end{equation} 
\indent The relations between EM fields $\textbf{E}'$, $\textbf{D}'$, $\textbf{B}'$, $\textbf{H}'$, and the unit wave vector  $\textbf{\^n}'$ in the medium-rest frame are shown in Fig.\,\ref{fig3}. Because $\mb{\Pi}'\cdot\textbf{E}'=0$ and $\mb{\Pi}'\cdot\textbf{H}'=0$ hold, we have $\mb{\Pi}'\parallel\textbf{\^n}'$, leading to $\delta\textbf{\^n}'\cdot\mb{\Pi}'=0$.  Inserting $\delta\textbf{\^n}'\cdot\mb{\Pi}'=0$ into Eq.\,(\ref{eq21}), we have $\Delta=0$.  Then inserting $\Delta=0$  into Eq.\,(\ref{eq11}), we have the classical group velocity \\
\begin{equation}
\textbf{v}_{\mathrm{gr-c}}=\frac{\textbf{S}}{W_{\mathrm{em}}} \hspace{2mm}
\left( \parbox{43mm}{for a \emph{moving} non-dispersive uniform medium} \right).
\label{eq23}
\end{equation} \vspace{0.5mm}

Obviously, Eq.\,(\ref{eq23}) holds in any inertial frames.  Especially, when observed in the frames with respect to which the medium moves parallel to the wave vector ($\mb{\beta}c\parallel\mathbf{k}_{\mathrm{w}}$), leading to $\textbf{S}_{\mathrm{pseu}}=0$ \cite{r11}, from Eq.\,(\ref{eq23}), and Eqs.\,(\ref{eq5}) and (\ref{eq6}), we have  \vspace{0.5mm}
\begin{equation}
\textbf{v}_{\mathrm{gr-c}}=\frac{\textbf{S}}{W_{\mathrm{em}}}=\textbf{v}_{\mathrm{ph}}.
\label{eq24}
\end{equation} 
However observed in the other frames, the dielectric parameters become tensors, leading to $\textbf{S}_{\mathrm{pseu}}\neq 0$, and we have 
\begin{equation}
\textbf{v}_{\mathrm{gr-c}}=\frac{\textbf{S}}{W_{\mathrm{em}}}\neq \textbf{v}_{\mathrm{ph}}, ~~ \textrm{with} ~~|\textbf{v}_{\mathrm{gr-c}}|>|\textbf{v}_{\mathrm{ph}}|.
\label{eq25}
\end{equation} %\vspace{0.1mm}

According to the principle of relativity, as mentioned before, the group velocity should be also equal to the phase velocity in a non-dispersive medium; however, from Eqs.\,(\ref{eq24}) and (\ref{eq25}) we find that the classical group velocity $\textbf{v}_{\mathrm{gr-c}}$ is equal to the phase velocity in some frames, while it is not in the others, which is not consistent with the principle of relativity.

In contrast, from Eq.\,(\ref{eq16}) and Eq.\,(\ref{eq6}) we have $\textbf{v}_{\mathrm{gr}}=\textbf{v}_{\mathrm{ph}}=\textbf{S}_{\mathrm{power}}/W_{\mathrm{em}}$ (group velocity = phase velocity = energy velocity) in all inertial frames for a moving non-dispersive isotropic uniform medium.  This result is completely consistent with the principle of relativity.
%\vspace{2mm}

%****Section 5*****
\section{Conclusions and remarks}
\label{s5}
In this paper, we used a plane wave to test the justification of the definition of classical group velocity.   We have shown both from the anisotropic \emph{symmetric stationary} medium and \emph{asymmetric moving} medium that, the classical group velocity $\textbf{v}_{\mathrm{gr-c}}=\partial\omega/\partial\textbf{k}_{\mathrm{w}}$  indeed has some flaws.  In contrast, the suggested modified definition $\textbf{v}_{\mathrm{gr}}=\textbf{\^n}\partial\omega/\partial |\textbf{k}_{\mathrm{w}}|$ has removed the flaws.  For the isotropic medium, the two definitions are equivalent.  %\vspace{2mm}
 
The differences between the classical and modified group velocities for a non-dispersive, lossless, non-conducting, anisotropic uniform medium are outlined below.
\begin{itemize}
\item  ~\emph{Classical group velocity} $\textbf{v}_{\mathrm{gr-c}}$.  For a symmetric stationary medium, $\textbf{v}_{\mathrm{gr-c}}=\textbf{S}/W_{\mathrm{em}}$ holds and it may be greater than the phase velocity, because the Poynting vector  $\textbf{S}$ contains a pseudo-power flow $\textbf{S}_{\mathrm{pseu}}$, which is never responsible for any EM power flowing at any places for any time.  For an asymmetric moving medium, $\textbf{v}_{\mathrm{gr-c}}=\textbf{S}/W_{\mathrm{em}}$  holds in all inertial frames, but $\textbf{v}_{\mathrm{gr-c}}=\textbf{v}_{\mathrm{ph}}$ holds in some frames while it does not in the others.

\item ~\emph{Modified group velocity} $\textbf{v}_{\mathrm{gr}}$. ~~ No matter for a stationary ~medium ~or ~for a moving medium, $\textbf{v}_{\mathrm{gr}}=\textbf{S}_{\mathrm{power}}/W_{\mathrm{em}}=\textbf{v}_{\mathrm{ph}}$ (group velocity = energy velocity = phase velocity) always holds, where $\textbf{S}_{\mathrm{power}}$  is the real power flow.
\end{itemize}  %\vspace{2mm}

The classical group velocity has two flaws:
\begin{itemize}
\item[(i)] ~\emph{Breaking Fermat's principle}.  Although the classical group velocity $\textbf{v}_{\mathrm{gr-c}}=\textbf{S}/W_{\mathrm{em}}$ has no contradiction with the energy conservation equation $ -\nabla\cdot ~\textbf{S}=\partial W_{\mathrm{em}}/\partial t$  (instead of energy conservation law, confer footnote \ref{f5}), the Poynting vector $\textbf{S}=\textbf{E}\times\textbf{H}$  may includes a pseudo-power flow $\textbf{S}_{\mathrm{pseu}}\ne 0$  which is perpendicular to the wave vector, resulting in  $\textbf{S}=\textbf{E}\times\textbf{H}$ \emph{not} parallel to the wave vector.  However Fermat's principle requires that the direction of EM energy transport must be parallel to the wave vector.  Thus in the classical group velocity $\textbf{v}_{\mathrm{gr-c}}=\textbf{S}/W_{\mathrm{em}}$, $\textbf{S}=\textbf{E}\times\textbf{H}$ as an EM power flow may violate Fermat's principle.  In other words, if the classical group velocity is taken as energy velocity for a plane wave in a lossless, non-dispersive, non-conducting anisotropic uniform medium, it is not consistent with the Fermat's principle.  [Note: $\textbf{S}_{\mathrm{pseu}}\ne 0$ results in both (a) the classical group velocity is greater than the phase velocity and (b) the Poynting vector as power flow violates Fermat's principle.]

\item[(ii)] \emph{Breaking the principle of relativity}.  For a moving isotropic medium, in the medium-rest frame the Poynting vector $\textbf{S}=\textbf{E}\times\textbf{H}$  is parallel to the wave vector, and as a power flow, it is consistent with the Fermat's principle and so is the classical group velocity  $\textbf{v}_{\mathrm{gr-c}}=\textbf{S}/W_{\mathrm{em}}$.  But in a general moving frame, the moving isotropic medium becomes anisotropic, and $\textbf{S}=\textbf{E}\times\textbf{H}$  is not parallel to the wave vector.  Thus as a power flow, $\textbf{S}=\textbf{E}\times\textbf{H}$  is not consistent with the Fermat's principle and neither is the classical group velocity  $\textbf{v}_{\mathrm{gr-c}}=\textbf{S}/W_{\mathrm{em}}$.  In other words, $\textbf{v}_{\mathrm{gr-c}}=\textbf{S}/W_{\mathrm{em}}$ satisfies the Fermat's principle in some inertial frames, while it does not in the others, resulting in the Fermat's principle not equally valid in all inertial frames, and thus breaking the principle of relativity.
\end{itemize}

As we have known, the classical group velocity denotes the energy velocity in an \emph{isotropic} medium while it may not in an \emph{anisotropic} medium.  From this it follows that the same classical group-velocity definition may have different physical implications, depending on the dielectric properties of medium, which, obviously, is a serious inconsistency theoretically.  Thus the modified definition proposed in the paper provides a solution to remove the inconsistency.

One might argue that there is no need to introduce the modified definition of group velocity if we agree that the group velocity in an anisotropic medium is not related to the transport velocity of signal energy.  That is exactly true.  However besides incurring the theoretical inconsistency mentioned above, such an agreement would contradict the well-accepted concept that group velocity is the velocity of EM energy transport, as indicated at the beginning of the paper.   

One might question the justification of the proposed modified definition.  In fact, for a plane wave in a lossless, non-dispersive, non-conducting uniform medium, the modified group velocity satisfies the most basic principles in physics, which is illustrated below.

\begin{itemize}
\item[(1)] The modified group velocity for the plane wave is given by $\textbf{v}_{\mathrm{gr}}=\textbf{S}_{\mathrm{power}}/W_{\mathrm{em}}$  (group velocity = energy velocity), where $\textbf{S}_{\mathrm{power}}$  and  $W_{\mathrm{em}}$ satisfies EM energy conservation equation $-\nabla\cdot\textbf{S}_{\mathrm{power}}=\partial W_{\mathrm{em}}/\partial t$ [confer Eq.\,(\ref{eq28})]. Thus the modified group velocity $\textbf{v}_{\mathrm{gr}}$  is consistent with EM energy conservation equation.  

\item[(2)] Fermat's principle requires that EM power propagate parallel to the wave vector while $\textbf{S}_{\mathrm{power}}$  is parallel to the wave vector, as shown in footnote \ref{f4}. Thus the modified group velocity $\textbf{v}_{\mathrm{gr}}=\textbf{S}_{\mathrm{power}}/W_{\mathrm{em}}$ satisfies Fermat's principle.  
 
\item[(3)] For a moving medium the modified group velocity has the same form and the same physical implication in all inertial frames, which is given by   $\textbf{v}_{\mathrm{gr}}=\textbf{S}_{\mathrm{power}}/W_{\mathrm{em}}=\textbf{V}_{\mathrm{ph}}$ (group velocity = energy velocity = phase velocity).  Thus the modified group velocity $\textbf{v}_{\mathrm{gr}}$ satisfies the principle of relativity. 
\end{itemize}

From above item (2) and item (3), we find that the modified group velocity has removed the flaws that the classical group velocity has.

We have known that required by Fermat's principle, the direction of the modified group velocity must be along the wave vector or phase velocity.   However one might ask: Why must its amplitude be the same in form as the one-dimensional case?  That is because the modified group velocity must satisfy 3D-to-1D correspondence principle.  In fact, we have verified that the modified group velocity is exactly equal to the rigidly-moving velocity of a light pulse in a lossless, non-dispersive, non-conducting, uniform medium (confer \ref{appa}).  

It is interesting to point out that the classical group velocity $\textbf{v}_{\mathrm{gr-c}}$ has its group-velocity four-vector \cite{r16} although it has some flaws that we have shown.  Below we would like to make some comments on the Lorentz property of a physical quantity and the principle of relativity.

Principle of relativity and constancy of the light speed in free space are two basic postulates in special theory of relativity.  The principle of relativity states that the laws of physics are the same in form in all inertial frames of reference \cite{r17}.  This principle is \emph{a restriction but also a guide} in formulating physical theories \cite{r11}.  

According to this principle, there is no preferred inertial frame for descriptions of physical phenomena; thus Maxwell equations, global momentum and energy conservation laws, Fermat’s principle, and Einstein light-quantum hypothesis are equally valid in any inertial frames, no matter whether the medium is moving or at rest, and no matter whether the space is fully or partially filled with a medium \cite{r11}.

A uniform plane electromagnetic wave in free space, which is a fundamental solution ~to Maxwell equations, propagates at the light speed ~in all directions.  Consequently, when directly applying this principle to Maxwell equations, one may find that the light speed must be the same in all inertial frames, in other words, the principle of relativity requires the constancy of light speed \cite{r18,r19}.\footnote
{\label{f6} \emph{Constancy of photon speed in free space.}  According to the principle of relativity, Einstein light-quantum hypothesis, momentum-energy conservation law, and Maxwell equations are equally valid in all inertial frames.  Thus as the carriers of light energy and momentum, any photons in free space keep moving \emph{uniformly} after they leave a source observed in any inertial frames.  On the other hand, observed far away from the source (especially at the infinity, which was used as an assumption to derive Doppler effect in the 1905 paper by Einstein \cite{r17}), the light wave behaves as a (local) plane wave, while the photons for a plane wave move \emph{at the light speed} in all inertial frames due to the invariance of Maxwell equations.  From this we conclude that the photons in free space, \emph{regardless of how created}, move at the light speed in all directions \emph{independently of the motion of the source or the observer}, which is the direct result from the principle of relativity.  In a recent experimental demonstration, Giovannini and coworkers claim that ``spatially structured'' photons travel in free space slower than the speed of light \cite{r4}; however, the authors do not provide an answer to the fundamental question of why the slower-than-light-speed photons in free space do not break the principle of relativity; thus calling into question their claim.
} 
  From this, the time-space coordinates constitute a Lorentz four-vector and the EM fields follow four-tensor Lorentz transformations, \emph{no matter whether the space is filled with a medium or not}.\footnote
{\label{f7} As shown in Sec.\,11.\,9 of the textbook by Jackson \cite{r15}, the general Maxwell equations are given by $[\nabla\times\mathbf{H}-\partial(c\mathbf{D})/\partial(ct),\nabla\cdot(c\mathbf{D})]=(\mathbf{J},c\rho)$ and $[\nabla\times\mathbf{E}-\partial(-c\mathbf{B})/\partial(ct),\nabla\cdot(-c\mathbf{B})]=(\mathbf{0},0)$, which can be written as $\partial_{\mu}G^{\mu\nu}(\mathbf{D},\mathbf{H})=J^{\nu}$  and $\partial_{\mu}\mathscr{F}^{\mu\nu}(\mathbf{B},\mathbf{E})=0$.  If $G^{\mu\nu}(\mathbf{D},\mathbf{H})$ and $\mathscr{F}^{\mu\nu}(\mathbf{B},\mathbf{E})$ are four-tensor Lorentz covariant, then the Maxwell equations must be the same in form in all inertial frames, and $(\mathbf{J},c\rho)$  must be a four-vector, leading to the Lorentz invariance of Planck constant, electron charge, and fine structure constant \cite{r11}.  $\mathscr{F}^{\mu\nu}$  is the dual field-strength tensor of $F^{\mu\nu}=\partial^{\mu}A^{\nu}-\partial^{\nu}A^{\mu}$.  Thus mathematically,   $F^{\mu\nu}$ can be defined from $\mathscr{F}^{\mu\nu}$  without introducing $A^{\mu}$.  Note that  in  $G^{\mu\nu}(\mathbf{D},\mathbf{H})$, with $\mathbf{D}$ replaced by $(-\mathbf{B})$ and $\mathbf{H}$ replaced by $\mathbf{E}$, we obtain $\mathscr{F}^{\mu\nu}(\mathbf{B},\mathbf{E})$, namely $\mathscr{F}^{\mu\nu}(\mathbf{B},\mathbf{E})=G^{\mu\nu}(-\mathbf{B},\mathbf{E})$.
}  However it is not required that every physical quantity be a Lorentz scalar, a four-vector, or a four-tensor ...; resulting in ``intrinsic Lorentz violation (ILV)'' \cite{r19}.  The simplest example is the phase velocity, which never can be used to constitute a Lorentz velocity four-vector, because the phase velocity $\textbf{v}_{\mathrm{ph}}=\textbf{\^n}(\omega/|\textbf{k}_{\mathrm{w}}|)$  is defined based on the wave four-vector $K^{\mu}=(\textbf{k}_{\mathrm{w}},\omega/c)$, with a constraint of $\textbf{v}_{\mathrm{ph}}\parallel\textbf{k}_{\mathrm{w}}$ \cite{r11}.  Like the phase velocity, the modified group velocity  $\textbf{v}_{\mathrm{gr}}=\textbf{\^n}(\partial\omega/\partial|\textbf{k}_{\mathrm{w}}|)$ can't either, and there is nothing contradicting the principle of relativity.

It should be noted that, the ILV is essentially different from the ``Lorentz violation (LV)'' presented in \cite{r20}.  The ILV takes place within the frame of the two postulates (namely the principle of relativity and constancy of light speed), and it is completely consistent with the special relativity.  In contrast, the LV \cite{r20} describes deviations from the two postulates; for example, there has been a controversy recently about whether there are deviations in the time dilation predicted by special relativity in experiments of high-energy ions \cite{r21,r22}.

In addition, we also would like to make some comments on the definition of power flow.  The EM energy conservation equation is given by \cite{r23} 
\begin{equation}
-\nabla\cdot ~\textbf{S}=\frac{\partial W_{\mathrm{em}}}{\partial t}+\textbf{J}\cdot\textbf{E},
\label{eq26}
\end{equation}  
\noindent which means that the energy flowing into a differential box is equal to the increase of EM energy in the box plus the work on charge done by the electric field. If there is no conducting current existing ($\textbf{J}=0$), Eq.\,(\ref{eq26}) becomes 
\begin{equation}
-\nabla\cdot ~\textbf{S}=\frac{\partial W_{\mathrm{em}}}{\partial t}.
\label{eq27}
\end{equation} 
\indent In principle, EM field solutions can be obtained by solving Maxwell equations associated with their boundary conditions without any ambiguity.  However there does be some ambiguity for the definition of power flow in terms of above Eq.\,(\ref{eq27}).  Traditionally, $\textbf{S}=\textbf{E}\times\textbf{H}$ is defined as the power flow.  However by adding a term with a zero divergence to $\textbf{S}$, Eq.\,(\ref{eq27}) will not be affected \cite[p.9]{r6}\cite{r23}.   For example, inserting $\textbf{S}=\textbf{S}_{\mathrm{power}}+\textbf{S}_{\mathrm{pseu}}$ into Eq.\,(\ref{eq27}), with $\nabla\cdot\textbf{S}_{\mathrm{pseu}}\equiv 0$ taken into account we have the same-form conservation equation \\
\begin{equation}
-\nabla\cdot ~\textbf{S}_{\mathrm{power}}=\frac{\partial W_{\mathrm{em}}}{\partial t}.
\label{eq28}
\end{equation} 
\noindent Thus we can re-define $\textbf{S}_{\mathrm{power}}$ as the power flow.  For an isotropic medium, $\textbf{S}_{\mathrm{pseu}}=0$  and $\textbf{S}=\textbf{S}_{\mathrm{power}}$, and this re-definition has no effect \cite{r14}.

From above analysis we can see that, from the viewpoint of EM energy conservation equation, $\textbf{S}$ and $\textbf{S}_{\mathrm{power}}$ have the equal right to be the power flow.  However  $\textbf{S}$, as being a power flow, may contradict Fermat's principle, while $\textbf{S}_{\mathrm{power}}$ does not.  From this perspective, it is justifiable to take $\textbf{S}_{\mathrm{power}}$ as the correct power flow in an anisotropic medium (crystal) \cite{r14}.

It should be noted that there is a significant difference between ``energy conservation law" and ``energy conservation equation". As indicated in Sec.\,\ref{s1}, the energy conservation law is a fundamental postulate in physics, while the energy conservation equation is a specific mathematical formulation of the law.  That an EM power flow satisfies energy conservation equation is a necessary condition for this power flow to satisfy energy conservation law, but it is not a sufficient condition, as shown in footnote \ref{f5}. In an anisotropic medium, Poynting vector $\textbf{S}$ and the (real) power flow $\textbf{S}_{\mathrm{power}}$ both satisfy energy conservation \emph{equation}, but the Poynting vector as EM power flow may break energy conservation \emph{law}, as indicated in Sec.\,\ref{s2}, where Fermat's principle is shown to be consistent with the energy conservation law because $\textbf{S}_{\mathrm{power}}=W_{\mathrm{em}}\textbf{v}_{\mathrm{ph}}$  has already carried all the EM energy $W_{\mathrm{em}}$ propagating at $\textbf{v}_{\mathrm{ph}}\parallel\mathbf{k}_{\mathrm{w}}$.

It might be interesting to point out that, if $\textbf{S}_{\mathrm{pseu}}\neq 0$  were to be of power flow, a ``superluminal power flow'' could be constructed, given by  $\textbf{S}_{>c}=\textbf{S}+a\textbf{S}_{\mathrm{pseu}}$, where $a$ is an arbitrary constant, with $\textbf{S}_{>c}=\textbf{S}$ for $a=0$  and $\textbf{S}_{>c}=\textbf{S}_{\mathrm{power}}$ for $a=-1$.  Obviously, $\textbf{S}_{>c}$ satisfies energy conservation Eq. (\ref{eq27}) due to $\nabla\cdot\textbf{S}\equiv \nabla\cdot\textbf{S}_{>c}$.  Since $\textbf{S}_{\mathrm{pseu}}\perp\textbf{S}_{\mathrm{power}}$ holds, we have 
\begin{equation}
\left| \frac{\textbf{S}_{>c}}{W_{\mathrm{em}}} \right|=\frac{1}{|W_{\mathrm{em}}|}\sqrt{ \textbf{S}_{\mathrm{power}}^2+(a+1)^2 \textbf{S}_{\mathrm{pseu}}^2 }.
\label{eq29}
\end{equation} \\
\noindent From this we have $|\textbf{S}_{>c}/W_{\mathrm{em}}|>c$ holding for $\textbf{S}_{\mathrm{pseu}}\neq 0$  by a proper choice of $a$-value \cite{r14}.  This, one again, clearly shows that the EM energy conservation equation cannot uniquely determine the EM power flow without the energy conservation law or Fermat's principle taken into account.

Finally, we also would like to make some comments on the definition of group velocity itself.  It is shown in \ref{appa} that, in a non-dispersive, lossless, non-conducting uniform medium (no matter whether it is isotropic or anisotropic), a plane-wave light pulse propagates rigidly at the modified group velocity that is \emph{equal to the energy velocity}.  Thus in such a case, the modified group velocity has a clear, precise physical meaning, and it is an observable (at least, theoretically).  However it should be indicated that the group velocity is not a strict observable \emph{in general}, because the physical meaning of the definition itself is ambiguous. For example, the group velocity can exceed the vacuum light speed $c$ in an anomalous dispersion medium (confer footnote \ref{f9}), while the transport velocity of energy of a light pulse cannot exceed $c$, otherwise Einstein causality will be broken when considering that the pulse energy is an observable (confer footnote \ref{f11}).  Thus in such a case, the group velocity does not have physical meaning. 

In conclusion, in this paper we have provided new insight into light propagation and light-matter interactions by examining the physical implications of group velocity, EM power flow, Poynting theorem, energy conservation law, and Fermat's principle.  We have shown: (i) energy conservation law and Fermat's principle are physical postulates independent of Maxwell equations (Sec.\,\ref{s2}); (ii) Poynting vector as EM power flow in an anisotropic medium may break both Fermat's principle and energy conservation law (footnotes \ref{f4} and \ref{f5}); (iii) the modified definition of group velocity has removed the flaws that the classic definition has. We have also analyzed several experimental observations  in previous research works, with significant new results obtained, as shown in footnotes \ref{f6}, \ref{f10}, \ref{f11}, and \ref{f12}.   

%\newpage
%****Appendix A*****
\appendix
\section{Derivation, application, and physical implication of the modified group velocity}
\label{appa}
First, let us take a look of how Eq.\,(\ref{eq14}) is derived.  According to Landau-Lifshitz approach \cite{r1}, the refractive-index vector is defined as $\mathbf{n}_{\mathrm{d}}=\textbf{k}_{\mathrm{w}}/(\omega/c)$ with $n_{\mathrm{d}}=|\textbf{n}_{\mathrm{d}}|$, and we have the dispersion equation $\textbf{k}_{\mathrm{w}}^2-(\textbf{n}_{\mathrm{d}}~\omega/c)^2 =0$.  Inserting $\textbf{k}_{\mathrm{w}}=\textbf{n}_{\mathrm{d}}(\omega/c)$ into Eq.\,(\ref{eq1}), we have $\textbf{B}c=\textbf{n}_{\mathrm{d}}\times\textbf{E}$ and  $\textbf{D}c=-\textbf{n}_{\mathrm{d}}\times\textbf{H}$, leading to 
\begin{equation*}
\check{\epsilon}\cdot\textbf{E}c^2+\textbf{n}_{\mathrm{d}}\times\left[~\check{\mu}^{-1}\cdot(\textbf{n}_{\mathrm{d}}\times\textbf{E})\right]=0, 
\end{equation*}
which is a system of linear equations for $(E_x,E_y,E_z)$, and where $\check{\epsilon}$ and $\check{\mu}$ are, respectively, the dielectric permittivity and permeability tensors, and $\check{\mu}^{-1}$ denotes the inverse tensor of $\check{\mu}$.  From this, we obtain the (eigen) Fresnel equation $F(n_{\mathrm{d}},\epsilon_{ij},\mu_{ij},\theta_{\mathrm{w}},\phi_{\mathrm{w}})=0$ \cite{r1}, or $n_{\mathrm{d}}=n_{\mathrm{d}}(\epsilon_{ij},\mu_{ij},\theta_{\mathrm{w}},\phi_{\mathrm{w}})$, where $\epsilon_{ij}$ and $\mu_{ij}$ are the tensor elements, and $\theta_{\mathrm{w}}$ and $\phi_{\mathrm{w}}$ are the wave-vector angles so that $k_{\mathrm{w}x}=|\textbf{k}_{\mathrm{w}}|\sin\theta_{\mathrm{w}}\cos\phi_{\mathrm{w}}$, $k_{\mathrm{w}y}=|\textbf{k}_{\mathrm{w}}|\sin\theta_{\mathrm{w}}\sin\phi_{\mathrm{w}}$, and $k_{\mathrm{w}z}=|\textbf{k}_{\mathrm{w}}|\cos\theta_{\mathrm{w}}$ hold in the wave-vector space.  Since $n_{\mathrm{d}}$  does not explicitly contain  $|\textbf{k}_{\mathrm{w}}|$, we have $(\partial n_{\mathrm{d}}/\partial |\textbf{k}_{\mathrm{w}}|)_{\mathrm{\,explicit}}=0$  (with $\epsilon_{ij}$, $\mu_{ij}$, $\theta_{\mathrm{w}}$, and $\phi_{\mathrm{w}}$ kept constant).  If there is any dispersion, $n_{\mathrm{d}}$ implicitly contains $\omega$  through the dielectric constant tensors.  Thus from $n_{\mathrm{d}}=n_{\mathrm{d}}(\epsilon_{ij},\mu_{ij},\theta_{\mathrm{w}},\phi_{\mathrm{w}})$  we have  
\begin{equation}
\frac{\partial n_{\mathrm{d}}}{\partial|\textbf{k}_{\mathrm{w}}|} \equiv \left. \frac{\partial n_{\mathrm{d}}}{\partial|\textbf{k}_{\mathrm{w}}|} \right|_{\theta_{\mathrm{w}},\phi_{\mathrm{w}}=const} =\frac{\partial n_{\mathrm{d}}}{\partial\omega} \frac{\partial\omega}{\partial|\textbf{k}_{\mathrm{w}}|},   
\label{eqA1}
\end{equation} 
where 
\begin{equation}
\frac{\partial n_{\mathrm{d}}}{\partial\omega}\equiv \left. \frac{\partial n_{\mathrm{d}}}{\partial\omega} \right|_{\theta_{\mathrm{w}},\phi_{\mathrm{w}}=const} =\sum_{i,j} \left( \frac{\partial n_{\mathrm{d}}}{\partial\epsilon_{ij}}\frac{\partial \epsilon_{ij}}{\partial\omega}+\frac{\partial n_{\mathrm{d}}}{\partial\mu_{ij}}\frac{\partial \mu_{ij}}{\partial\omega}  \right).  
\label{eqA2}
\end{equation}  

From the dispersion equation  $|\textbf{k}_{\mathrm{w}}|^2-(n_{\mathrm{d}}~\omega/c)^2=0$, we have 
\begin{equation}
|\textbf{k}_{\mathrm{w}}|-n_{\mathrm{d}}\frac{\omega}{c}\left( \frac{\omega}{c}\frac{\partial n_{\mathrm{d}}}{\partial|\textbf{k}_{\mathrm{w}}|}+\frac{n_{\mathrm{d}}}{c}\frac{\partial\omega}{\partial|\textbf{k}_{\mathrm{w}}|} \right)=0.  
\label{eqA3}
\end{equation} 
Inserting Eq.\,(\ref{eqA1}) into Eq.\,(\ref{eqA3}), and from the modified definition $\textbf{v}_{\mathrm{gr}}=\textbf{\^n}\partial\omega/\partial|\textbf{k}_{\mathrm{w}}|$ with $\textbf{v}_{\mathrm{ph}}=\textbf{\^n}(\omega/|\textbf{k}_{\mathrm{w}}|)$  taken into account, we obtain Eq.\,(\ref{eq14}).

Now let us take a look of the application of Eq.\,(\ref{eq14}).  The modified group velocity formula Eq.\,(\ref{eq14}) has exactly the same form as that in the isotropic medium \cite{r15}.  However because of the anisotropy of $n_{\mathrm{d}}$, special attention is needed to the  $\partial/\partial\omega$-operation in Eq.\,(\ref{eq14}).  To better understand this, a specific example is given below.

For a plane wave in an electro-anisotropic uniaxial medium, which is the simplest anisotropic medium, we have  $\textbf{B}=\mu\textbf{H}$ and $\textbf{D}=\check{\epsilon}\cdot\textbf{E}$, where $\mu$ is the permeability constant with $\partial\mu/\partial\omega=0$ assumed, and $\check{\epsilon}$  is the permittivity tensor with its element matrix $(\epsilon_{ij})=\mathrm{diag}(\epsilon,\epsilon,\epsilon_z)$, which defines the $z$-axis as the optic axis.  From this we have the Fresnel's equation for the extraordinary wave, given by \cite[p.340]{r1}\,\cite[p.343]{r5} 
\begin{equation}
\frac{\cos^2\alpha_z}{\epsilon\mu}+\frac{\sin^2\alpha_z}{\epsilon_z\mu}=\left(\frac{c}{n_{\mathrm{d}}}\right)^2~~~\Rightarrow~~~n_{\mathrm{d}}=\frac{c\sqrt{\mu\epsilon_z}}{\sqrt{\sin^2\alpha_z+(\epsilon_z/\epsilon)\cos^2\alpha_z}},
\label{eqA4}
\end{equation}  
where  $\sin^2\alpha_z=\cos^2\alpha_x+\cos^2\alpha_y$, with $\alpha_x$,  $\alpha_y$, and  $\alpha_z$ being, respectively, the angles made by the unit wave vector ~$\textbf{\^n}$ ~with ~the ~$x$-, ~$y$-, ~and ~$z$-axes, ~namely  $\cos\alpha_x=k_{\mathrm{w}x}/|\textbf{k}_{\mathrm{w}}|=\sin\theta_{\mathrm{w}}\cos\phi_{\mathrm{w}}$,  $\cos\alpha_y=k_{\mathrm{w}y}/|\textbf{k}_{\mathrm{w}}|=\sin\theta_{\mathrm{w}}\sin\phi_{\mathrm{w}}$, and  $\cos\alpha_z=k_{\mathrm{w}z}/|\textbf{k}_{\mathrm{w}}|=\cos\theta_{\mathrm{w}}$.  From Eq.\,(\ref{eqA2}) and Eq.\,(\ref{eqA4}), with  $\alpha_x$,  $\alpha_y$, and  $\alpha_z$ kept constant we have 
\begin{equation}
\frac{\partial n_{\mathrm{d}}}{\partial\omega}=\frac{n_{\mathrm{d}}}{2c^2\mu\epsilon_z}\left[c^2\mu\frac{\partial\epsilon_z}{\partial\omega}-n_{\mathrm{d}}^2\cos^2\alpha_z\frac{\partial}{\partial\omega}\left(\frac{\epsilon_z}{\epsilon}\right)\right].
\label{eqA5}
\end{equation} 
%\newpage
Inserting Eqs.\,(\ref{eqA4}), (\ref{eqA5}), and $\textbf{v}_{\mathrm{ph}}=\textbf{\^n}(\omega/|\textbf{k}_{\mathrm{w}}|)=\textbf{\^n}(\omega/|\omega|)(c/n_{\mathrm{d}})$ into Eq.\,(\ref{eq14}), we can obtain the group velocity expression for the uniaxial medium, where the frequency sign $(\omega/|\omega|)=\pm$ denotes two possible waves propagating in opposite directions.

Note:  $\partial/\partial\omega$ only operates on the dielectric parameters of the right-hand side of Eq.\,(\ref{eqA4}),  $\epsilon_z$ and $\epsilon$, with the assumption $\partial\mu/\partial\omega=0$ taken into account. 

By setting  $\epsilon_z=\epsilon$, the anisotropy included in Eq.\,(\ref{eqA4}) and Eq.\,(\ref{eqA5}) disappears, leading to $n_{\mathrm{d}}=c(\mu\epsilon)^{1/2}$  and  $\partial n_{\mathrm{d}}/\partial\omega=[n_{\mathrm{d}}/(2\epsilon)](\partial\epsilon/\partial\omega)$, and Eq.\,(\ref{eq14}) is restored to the group velocity in an isotropic medium \cite{r15}, given by 
\begin{equation}
\textbf{v}_{\mathrm{gr}}=\frac{\omega}{|\omega|}\frac{c\textbf{\^n}}{c\sqrt{\mu\epsilon}\left(1+\displaystyle\frac{\omega}{2\epsilon}\displaystyle\frac{\partial\epsilon}{\partial\omega}\right)}.
\label{eqA6}
\end{equation} 
\indent Finally, let us take a look of the physical implications of modified group velocity.  As mentioned previously, the modified group velocity Eq.\,(\ref{eq14}) is the same in form as that in an isotropic medium \cite{r15}, except that $n_{\mathrm{d}}$  is anisotropic, depending on the direction of $\textbf{\^n}$.  However $\textbf{v}_{\mathrm{gr}}$ itself is defined in the  $\textbf{\^n}$-direction and the direction of $\textbf{\^n}$  is independent of $\omega$  and $|\textbf{k}_{\mathrm{w}}|$.

As shown in the textbook \cite{r15}, for a non-dispersive, lossless, non-conducting, \emph{isotropic} uniform medium, it is exactly true for the classical (= modified) group velocity to be equal to the\emph{ rigidly-moving velocity} of a plane-wave light pulse.  We will show below that it is also true for the modified group velocity in an \emph{anisotropic} medium.

Suppose that a plane-wave light pulse (wave train) propagating in the anisotropic medium has a frequency bandwidth of $\Delta\omega$  and a wave-number width of $\Delta|\textbf{k}_w|$, and the frequency bandwidth $\Delta\omega$  can be expanded into a series of  $\Delta|\textbf{k}_{\mathrm{w}}|$, given by 
\begin{equation}
\Delta\omega=\left( \textbf{\^n}\frac{\partial\omega}{\partial|\textbf{k}_{\mathrm{w}}|}\right)\cdot\textbf{\^n}(\Delta|\textbf{k}_{\mathrm{w}}|)+\left(\parbox{16mm}{high-order terms of $\Delta|\textbf{k}_{\mathrm{w}}|$}   \right),
\label{eqA7}
\end{equation} 
where $\textbf{\^n}(\partial\omega/\partial|\textbf{k}_{\mathrm{w}}|)$  is the modified group velocity $\textbf{v}_{\mathrm{gr}}$.  Thus the traditional interpretation for the group velocity \cite{r15} is applicable here: the modified group velocity $\textbf{v}_{\mathrm{gr}}$ is the velocity at which the light pulse travels along \emph{undistorted in shape} apart from an overall phase factor when $\Delta|\textbf{k}_{\mathrm{w}}|$  is so small that the effect of high-order terms of $\Delta|\textbf{k}_{\mathrm{w}}|$  can be \emph{ignored}.  Of course, if there are \emph{no} high-order terms in Eq.\,(\ref{eqA7}), the pulse travels exactly at $\textbf{v}_{\mathrm{gr}}$ without any distortions in shape, namely \emph{in a rigid way}.

What is the energy velocity for a finite-size plane-wave light pulse?  If the light pulse propagates exactly \emph{in a rigid way} at a velocity, then this velocity, without question, can be defined as the ``whole'' energy velocity of the pulse.  However because of the existence of dispersion in a practical medium, the pulse will be distorted more or less.  Thus the energy velocity is usually defined by the power flow divided by the EM energy density.  In fact, this is the ``local'' energy velocity, ignoring whether the pulse is distorted or not during its propagation ahead.  Fortunately, the ``local'' and ``whole'' energy velocities and the modified group velocity are equal in a non-dispersive medium ($\partial n_{\mathrm{d}}/\partial\omega=0$), because $\partial\omega/\partial|\textbf{k}_{\mathrm{w}}|=\omega/|\textbf{k}_{\mathrm{w}}|$  holds according to Eq.\,(\ref{eq14}), resulting in the holding of $\partial^{(n)}\omega/\partial|\textbf{k}_{\mathrm{w}}|^{n}=0$  for $n\geq 2$  and the \emph{disappearance} of all the high-order terms in Eq.\,(\ref{eqA7}), and further, the whole pulse moving at $\textbf{v}_{\mathrm{gr}}$ rigidly.

From above analysis we know that, for a non-dispersive, lossless, non-conducting, \emph{anisotropic} uniform medium, the modified group velocity Eq.\,(\ref{eq14}), which is equal to the phase velocity, is also equal to the \emph{rigidly-moving velocity} of a plane-wave light pulse.  Thus the modified ~definition ~is ~exactly ~accurate ~for ~a ~non-dispersive medium, and it can be taken as an approximate description of the energy transport velocity in a weak-dispersion medium.  However in strong-dispersion regions of a dispersion medium,\footnote
{\label{f8}  In a dispersion medium, there are the region of normal dispersion and the region of anomalous dispersion.  The strong dispersion regions include both the strong normal and anomalous dispersion regions \cite[p.310]{r15}.
} 
such an approximation may not valid at all; for example, the group velocity may exceed the vacuum light speed, which has been demonstrated by experiments in the region of strong anomalous dispersion \cite{r24}, and in such a case the group velocity itself is not a meaningful physical quantity \cite{r15,r25,r26}.\footnote
{\label{f9}  Traditionally, it has been thought that ``group velocity is \emph{generally} not a useful concept in regions of anomalous dispersion'' \cite{r15}.  But Milonni disagrees, criticizing that ``group velocity ceases to have physical significance in the case of anomalous dispersion, when it can exceed $c$''  is a long-standing misconception, because the group velocity in such a case still retains ``its meaning as the velocity of nearly \emph{undistorted} pulse propagation, as experiments have shown'' \cite{r27}.   However if superluminal group velocity has physical meaning and trains of nearly \emph{undistorted} superluminal pulses are physical, then according to principles of digital telecommunications, the trains of pulses can be used to constitute signals to propagate superluminally, which possibly results in the contradiction with special relativity.  
}$^{,}$\footnote
{\label{f10}  \emph{Will a causality violation not happen when light pulses pass through material media faster than the speed of light?}  To explain why a light pulse propagating at the superluminal group velocity does not violate Einstein causality, Boyd and Gauthier argue that ``no information is carried by the smooth portions of the waveform'', and the information content is contained in the pulse front of ``discontinuity'', while the front propagates at the speed of light in a vacuum $c$; as a result, the information velocity is always equal to $c$ (namely the invariance of information velocity), although the group velocity can take on any value \cite{r28}. However this argument is not consistent with Maxwell EM theory if considering that the information transfer is a physical process. As we know, Maxwell equations are a system of first-order partial differential equations, and all EM fields have first-order partial derivatives unless on the EM boundaries.  Thus all EM fields are differentiable, and they must be of continuous functions; and so must the EM fields for a physical light pulse that is a solution of Maxwell equations.  From this we conclude that a physical light pulse does not have any points of ``discontinuity''; thus calling into question Boyd-Gauthier argument that ``new information is encoded at each discontinuity'', and further the invariance of information velocity.  In other words, Boyd and Gauthier do not provide an answer to the fundamental question of why \emph{superluminal light pulses} do not violate Einstein causality.
}$^{,}$\footnote
{\label{f11}  \emph{Superluminal speed of energy transport?} In their superluminal light pulse experiments \cite{r24,r29}, Wang and coworkers claim that the probe light pulse is placed in the middle of two gain lines spectrally and it contains no spectral components to be amplified; thus the superluminal light pulse propagation is a result of the rephasing and interference of different-frequency component waves in the anomalous dispersion region that is lossless.  The authors argue that the existence of the lossless anomalous dispersion region is a result of the Kramers-Kronig relation which itself is based on the causality requirements of electromagnetic responses; from this they draw a conclusion that ``the observed superluminal light pulse propagation is not at odds with causality or special relativity''.  However, this conclusion does not seem sufficiently convincing if it is reviewed from the perspective of energy transport. As shown in Fig.\ 4 of their Letter \cite{r24}, the whole pulse \emph{intensity} profile observed is advanced by 62 ns nearly without any distortions.  The pulse energy, measured at the end of the anomalous dispersion medium ($z = L$), is equal to the integral of intensity $I(L,t)$ over time $t$, namely $\int I(L,t)\mathrm{d}t$ (confer Eq.\ (18) of Ref.\ \cite{r29}), and thus the light pulse energy is also advanced by 62 ns.  From this we judge that the light pulse energy must be transported faster than the speed of light in a vacuum (superluminally). However, Wang and coworkers do not provide an answer to the fundamental question of why this \emph{superluminal energy transport} does not violate special relativity; thus calling into question the conclusion that ``the observed superluminal light pulse propagation is not at odds with causality or special relativity''.
}$^{,}$\footnote
{\label{f12}  \emph{Superluminal communications?} In an experimental demonstration of superluminal light propagation based on Brillouin lasing oscillation, Zhang and coworkers claim that in the anomalous dispersion medium, the group velocity of the signal pulse does exceed the speed of light in a vacuum, and argue that this phenomenon ``provides a new way of opening up superluminal communications via optical fibers'' \cite{r30}; however, the authors do not provide an answer to the fundamental question of why \emph{superluminal communications} do not violate Einstein causality.
}$^{,}$\footnote
{\label{f13}  Recently, experimental observations in microwave regime indicate that the definition of group velocity is ``physically meaningless in the anomalously dispersive region'' \cite{r31}, supporting the conclusion obtained in the present paper that the group velocity is not a strict observable \emph{in general}, because the physical meaning of the definition itself is ambiguous.
}  \\

%\section*{References}

\newpage
\onecolumn
\begin{figure} % figure M1
\includegraphics[trim=1.0in 1.0in 1.0in 1.0in, clip=true,scale=1.0]{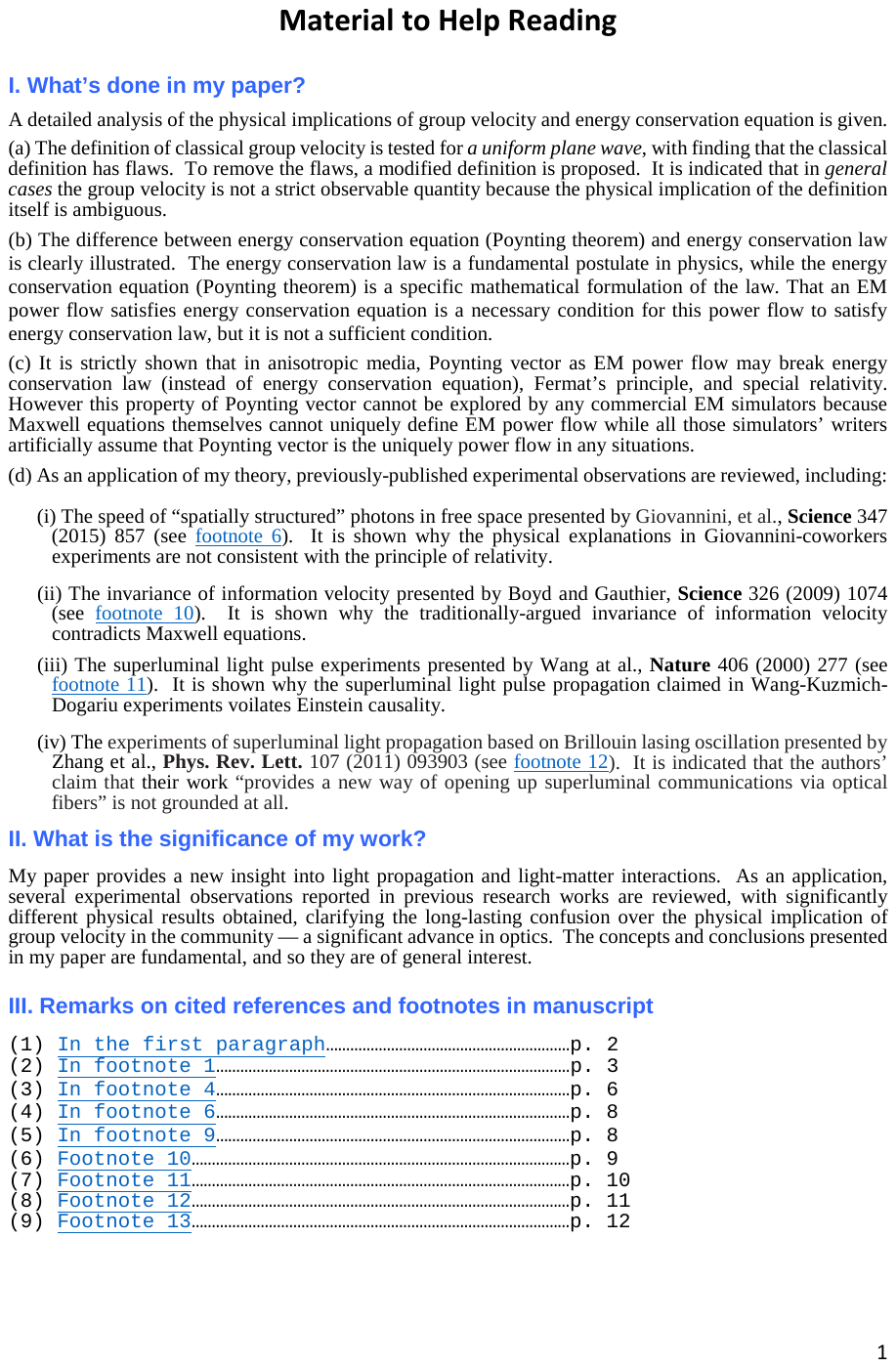}
%%\caption{}
\label{figM1}
\end{figure} 

\begin{figure} % figure M2
\includegraphics[trim=1.0in 1.0in 1.0in 1.0in, clip=true,scale=1.0]{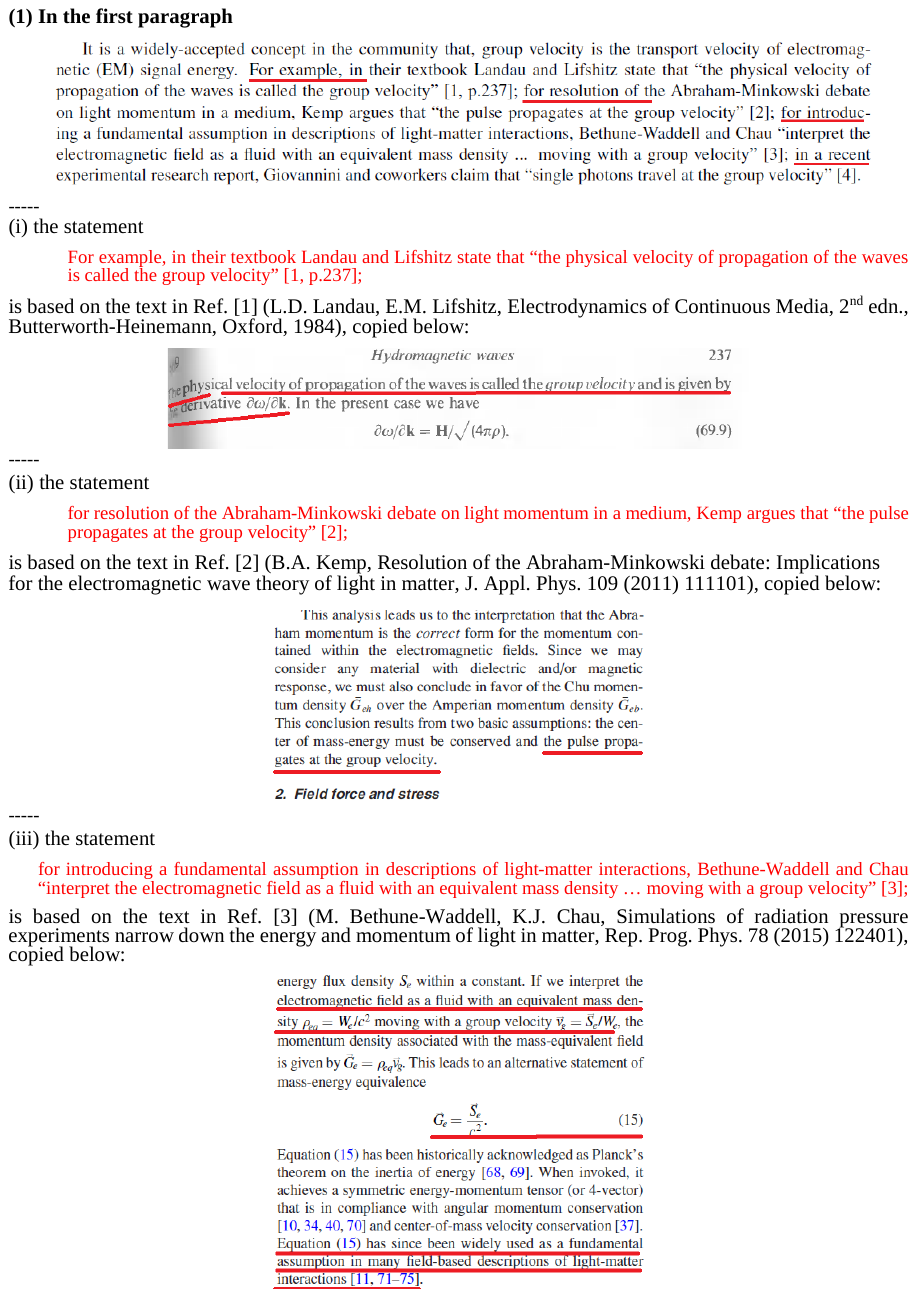}
%%\caption{}
\label{figM2}
\end{figure} 

\begin{figure} % figure M3
\includegraphics[trim=1.0in 1.0in 1.0in 1.0in, clip=true,scale=1.0]{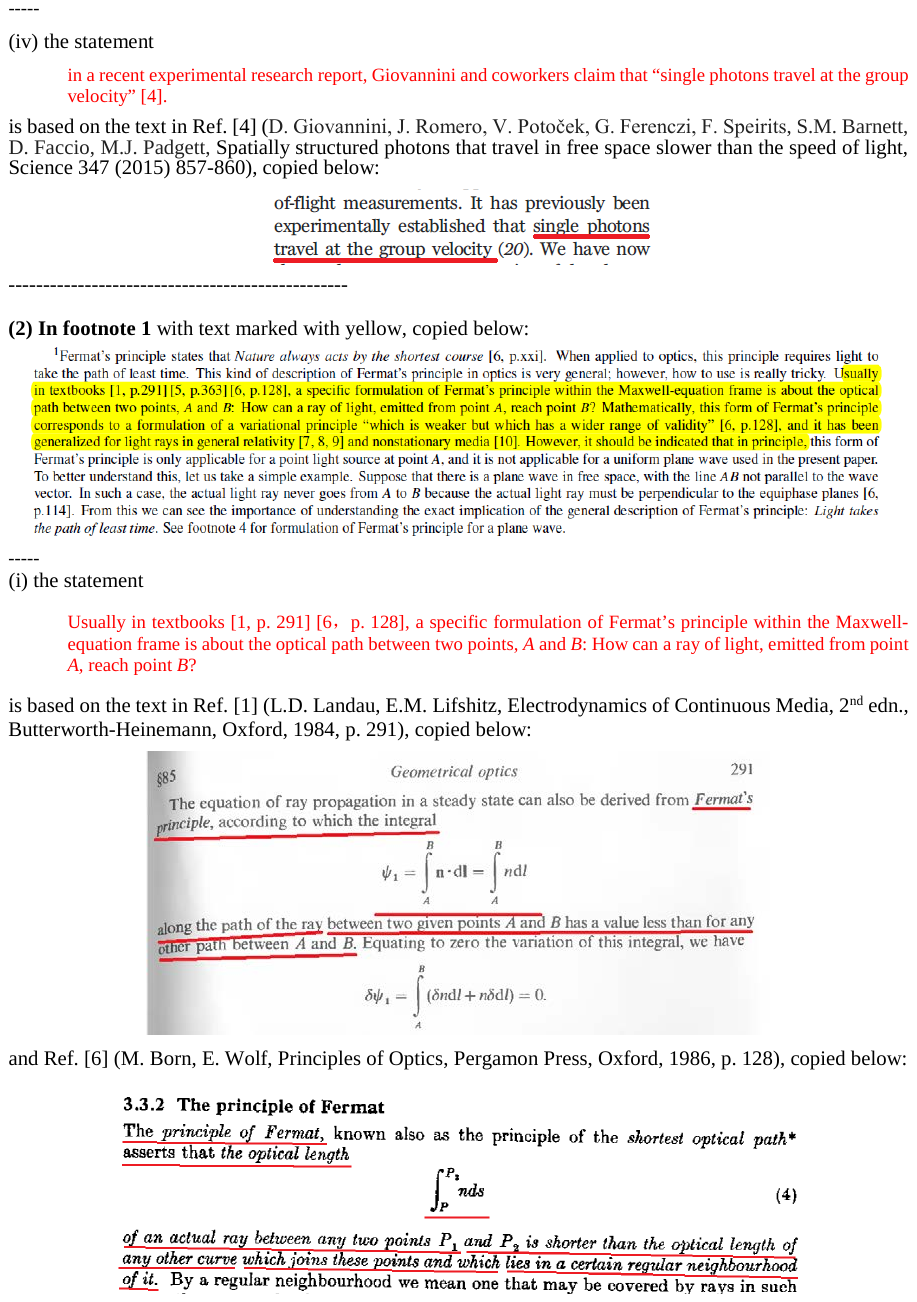}
%%\caption{}
\label{figM3}
\end{figure} 

\begin{figure} % figure M4
\includegraphics[trim=1.0in 1.0in 1.0in 1.0in, clip=true,scale=1.0]{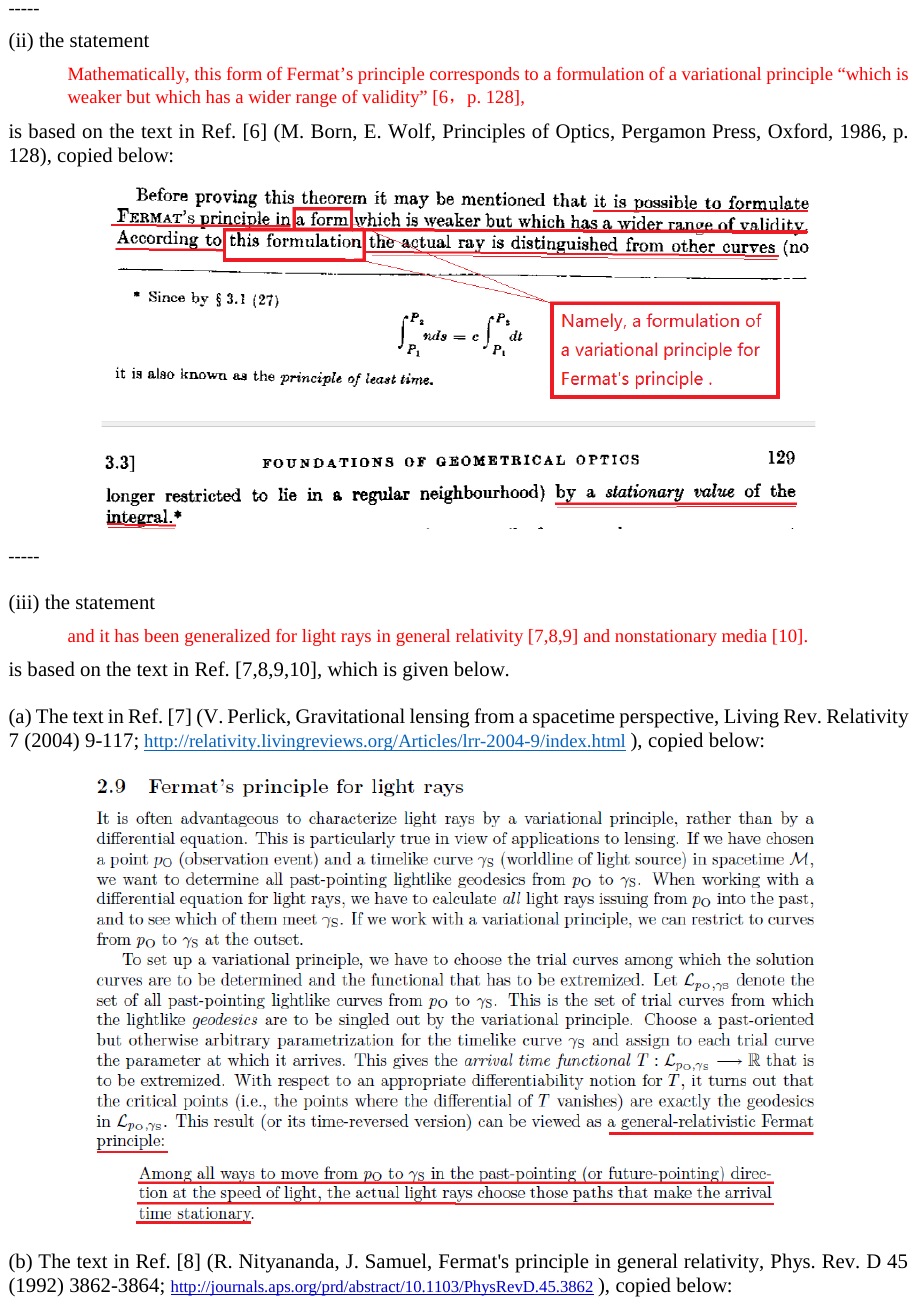}
%%\caption{}
\label{figM4}
\end{figure} 

\begin{figure} % figure M5
\includegraphics[trim=1.0in 1.0in 1.0in 1.0in, clip=true,scale=1.0]{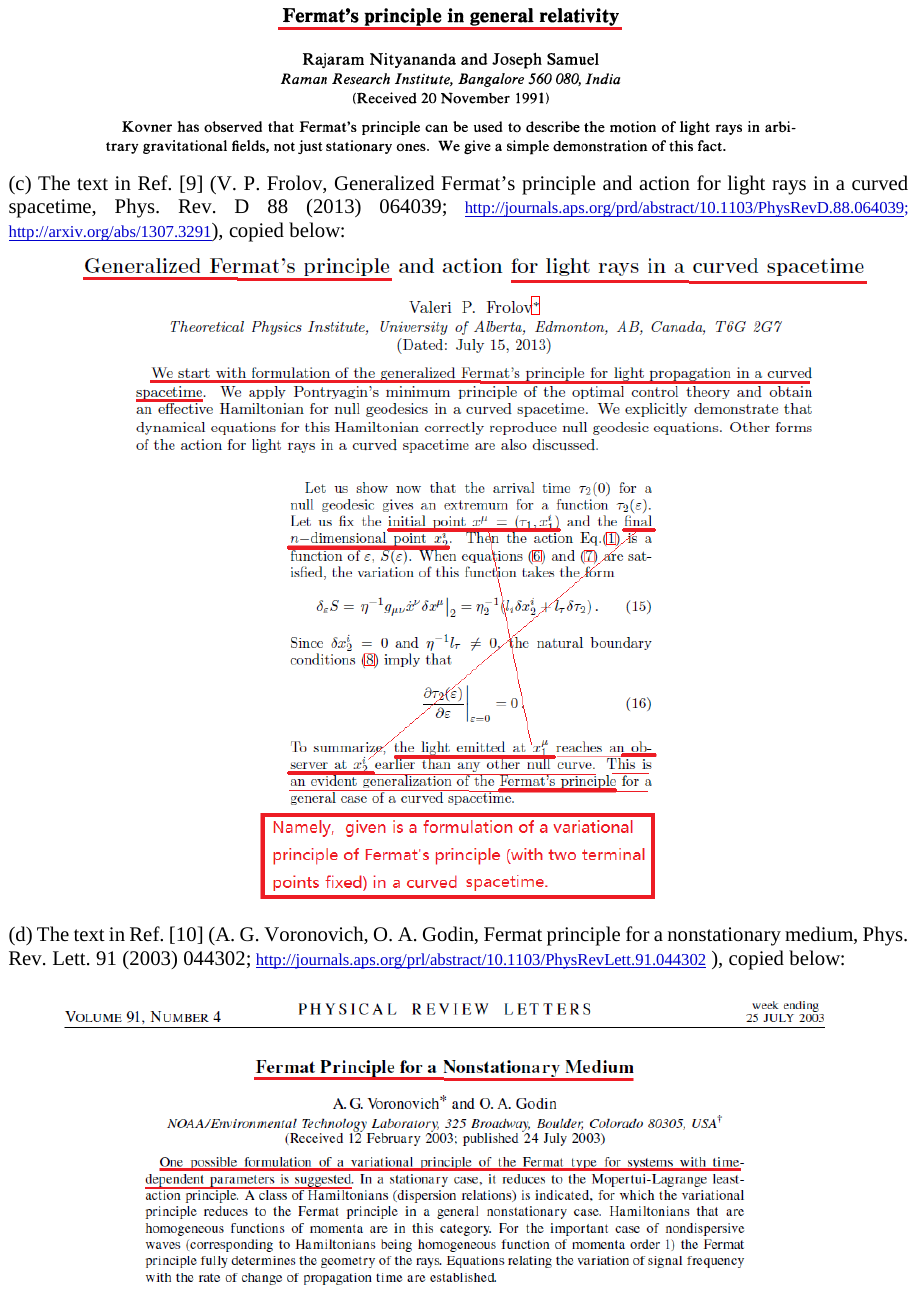}
%%\caption{}
\label{figM5}
\end{figure} 

\begin{figure} % figure M6
\includegraphics[trim=1.0in 1.0in 1.0in 1.0in, clip=true,scale=1.0]{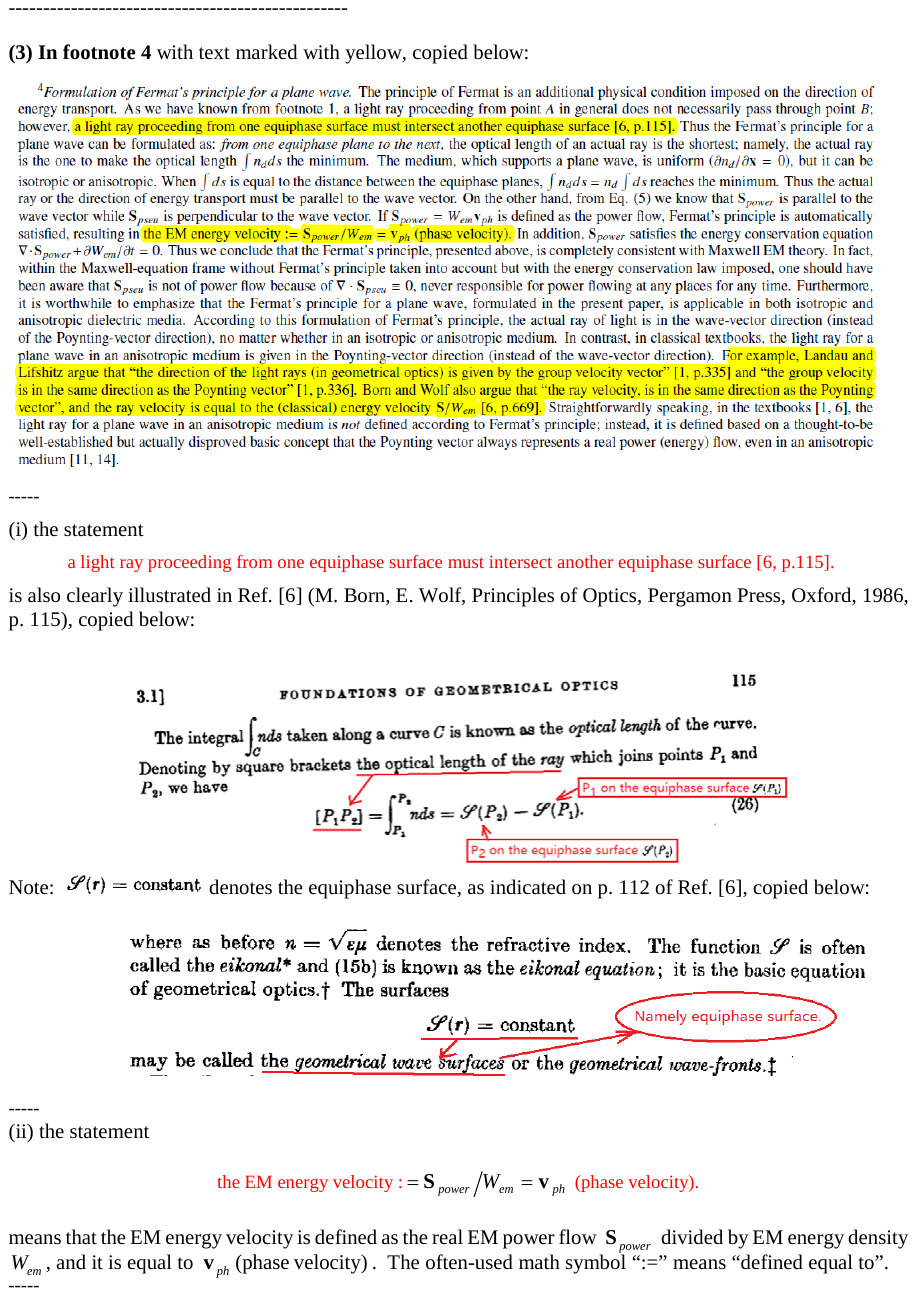}
%%\caption{}
\label{figM6}
\end{figure} 

\begin{figure} % figure M7
\includegraphics[trim=1.0in 1.0in 1.0in 1.0in, clip=true,scale=1.0]{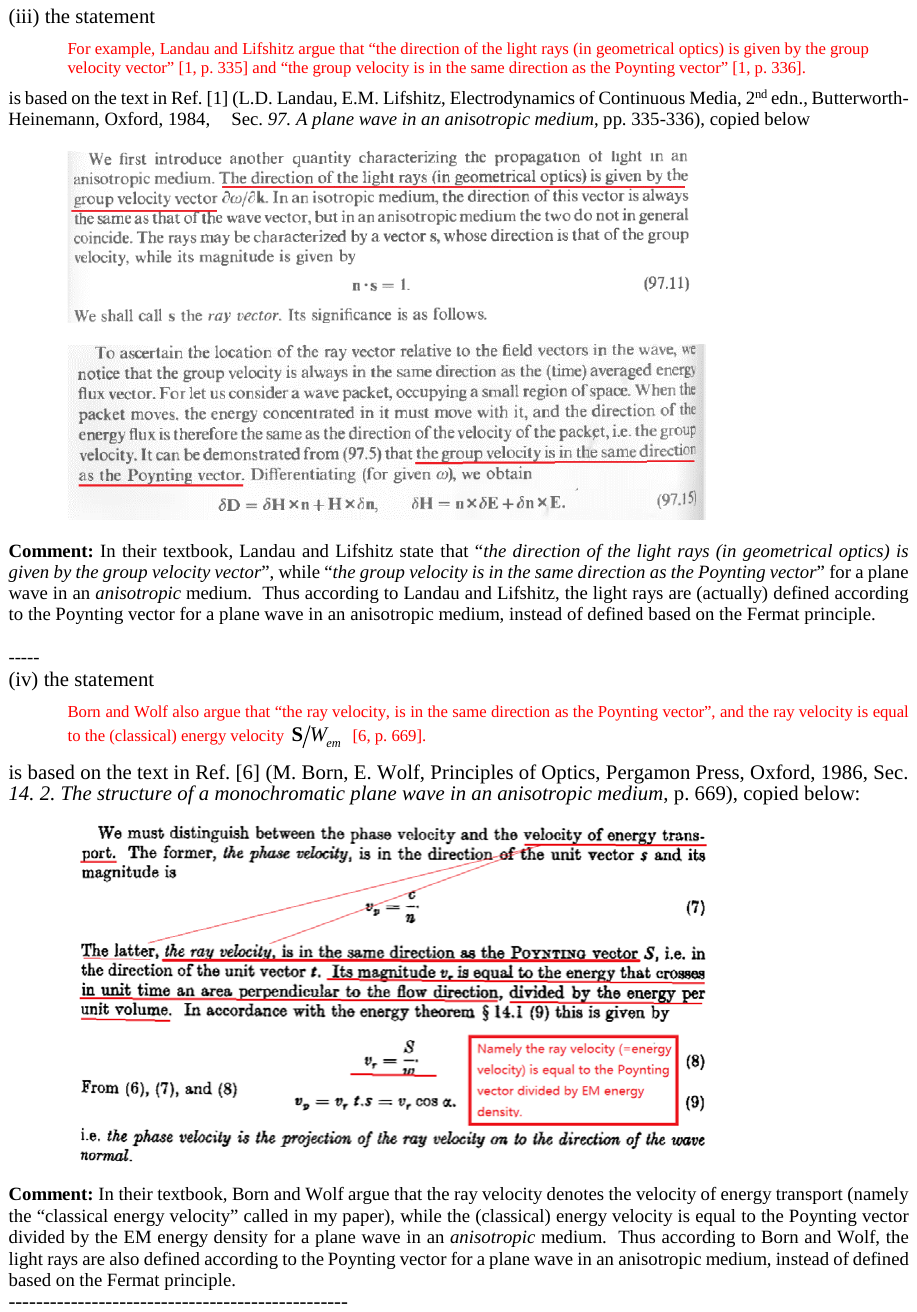}
%%\caption{}
\label{figM7}
\end{figure} 

\begin{figure} % figure M8
\includegraphics[trim=1.0in 1.0in 1.0in 1.0in, clip=true,scale=1.0]{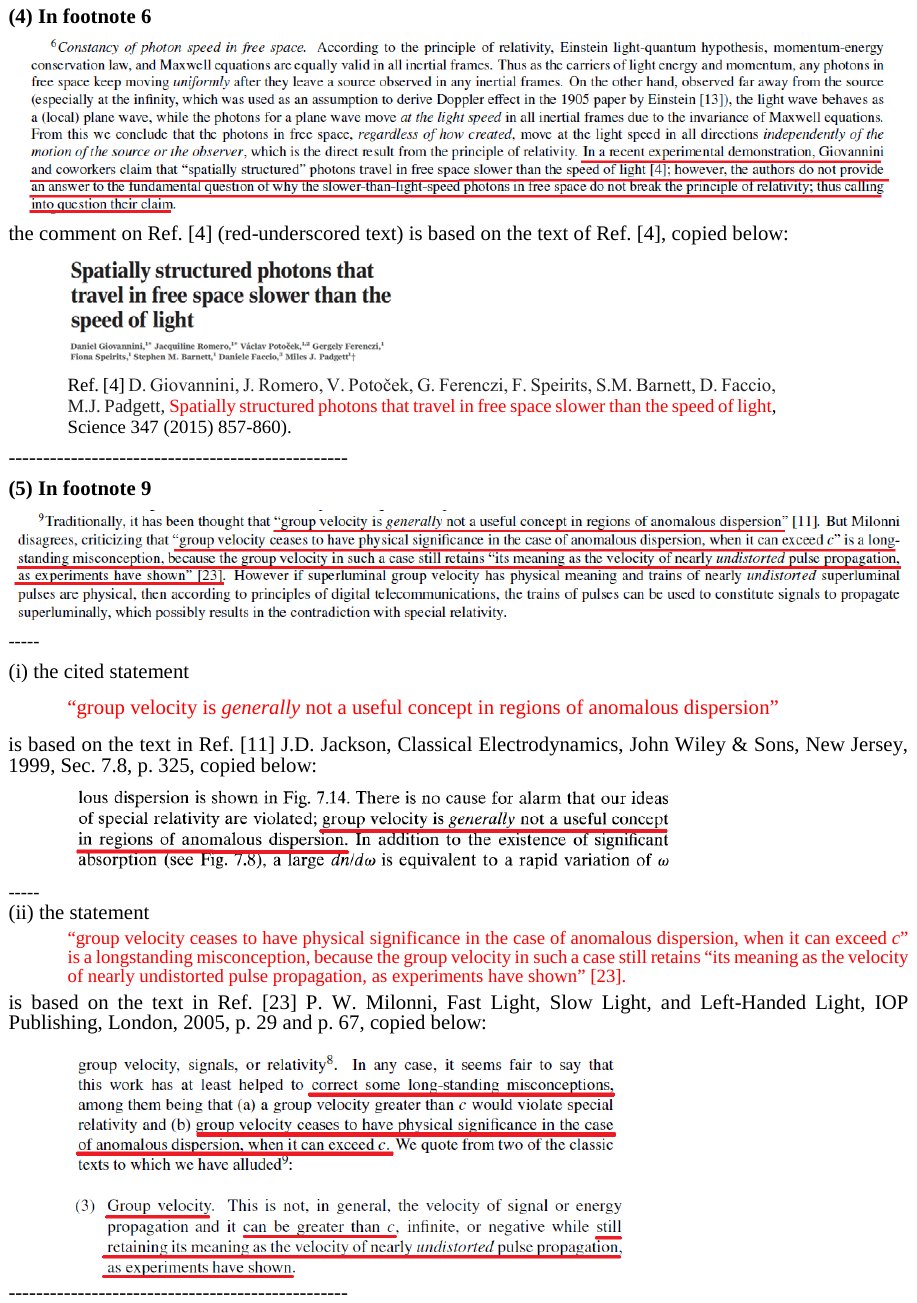}
%%\caption{}
\label{figM8}
\end{figure} 

\begin{figure} % figure M9
\includegraphics[trim=1.0in 1.0in 1.0in 1.0in, clip=true,scale=1.0]{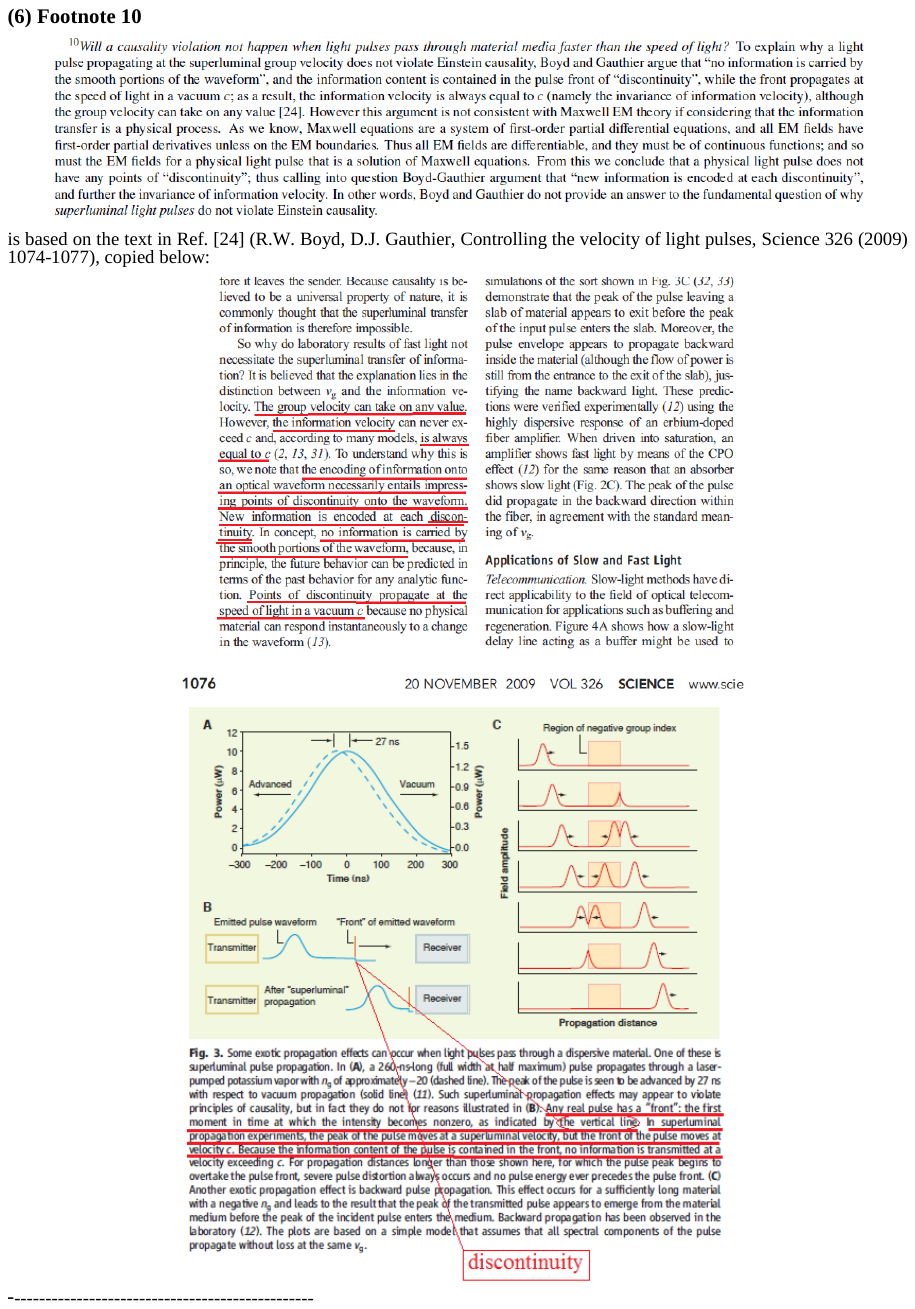}
%%\caption{}
\label{figM9}
\end{figure} 

\begin{figure} % figure M10
\includegraphics[trim=1.0in 1.0in 1.0in 1.0in, clip=true,scale=1.0]{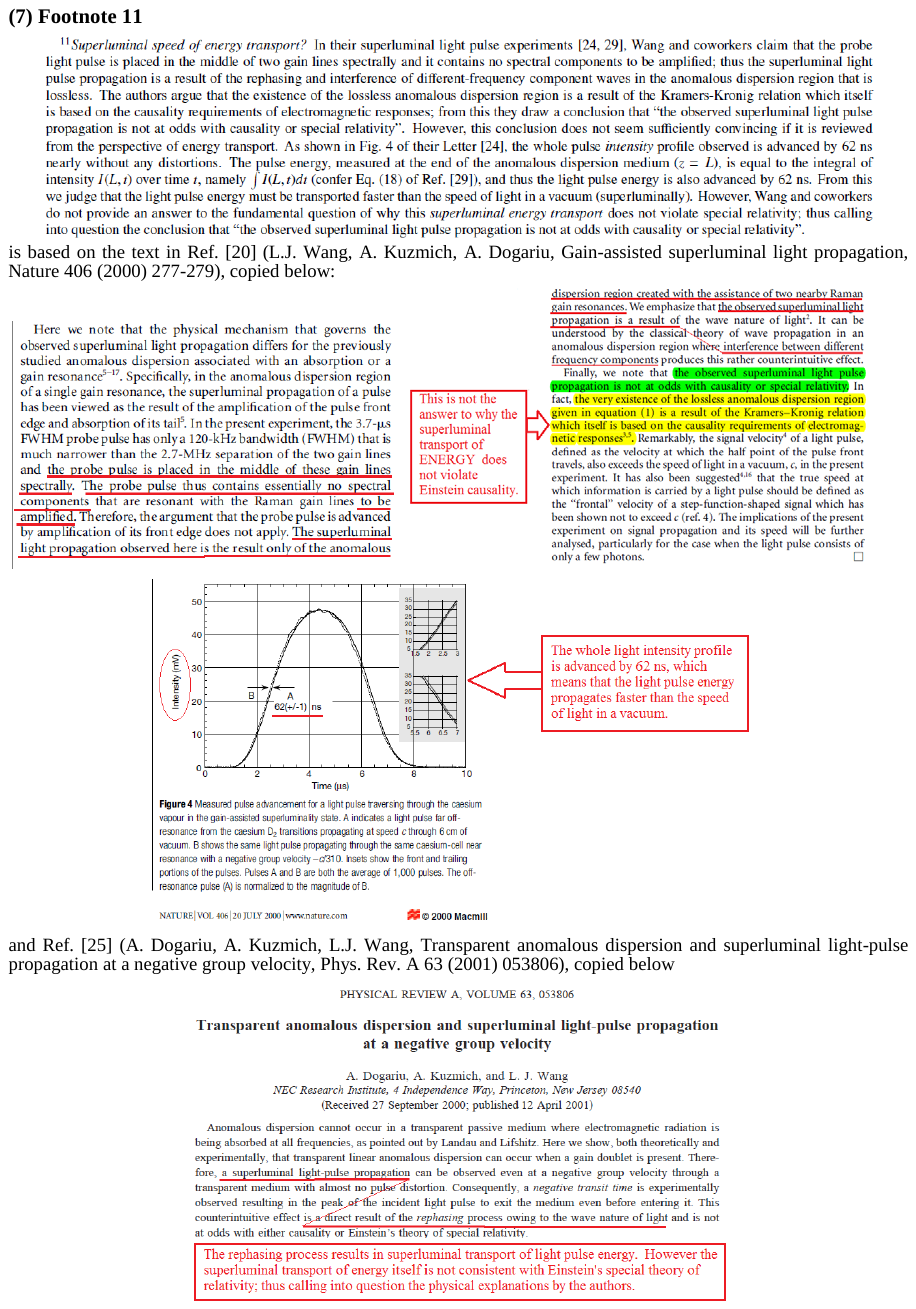}
%%\caption{}
\label{figM10}
\end{figure} 

\begin{figure} % figure M11
\includegraphics[trim=1.0in 1.0in 1.0in 1.0in, clip=true,scale=1.0]{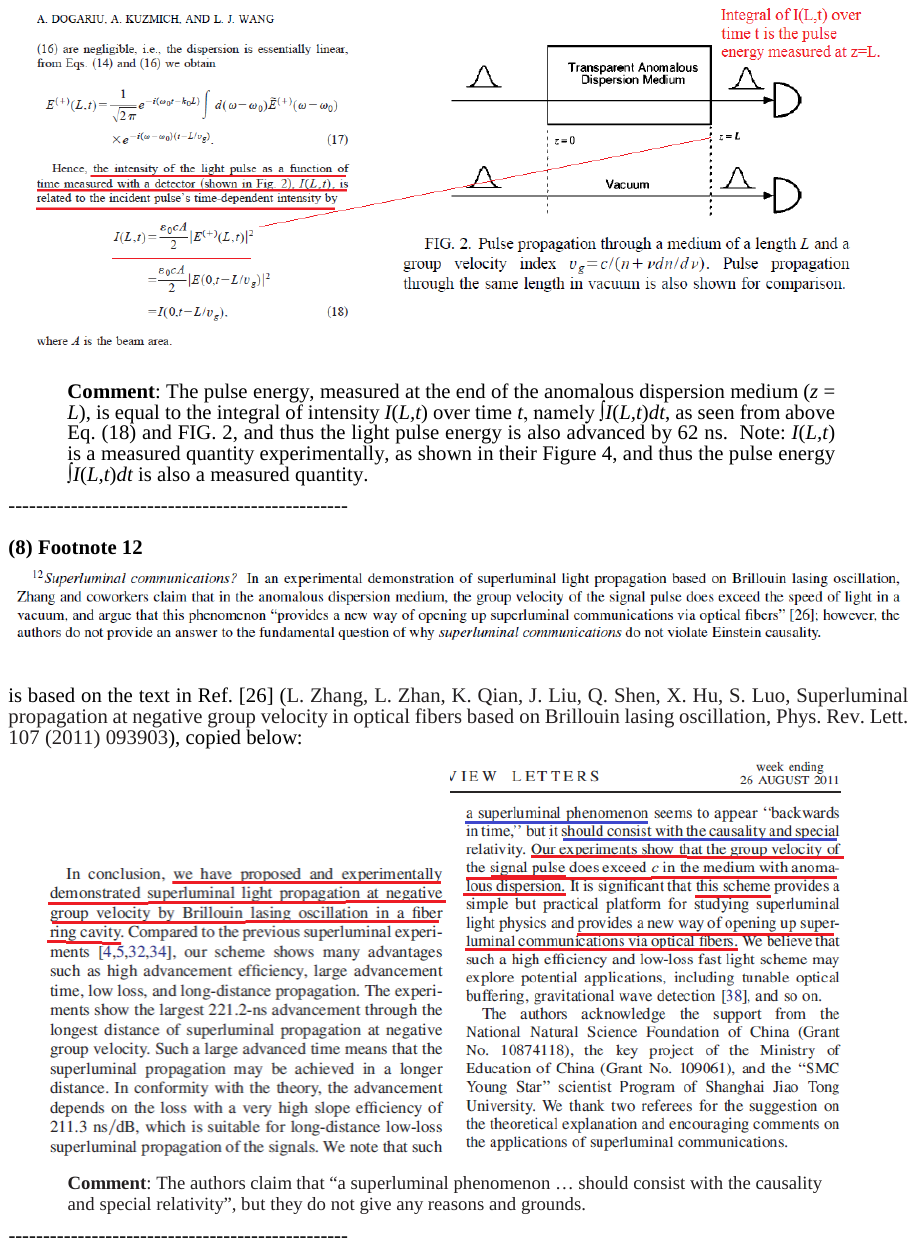}
%%\caption{}
\label{figM11}
\end{figure} 

\begin{figure} % figure M12
\includegraphics[trim=1.0in 1.0in 1.0in 1.0in, clip=true,scale=1.0]{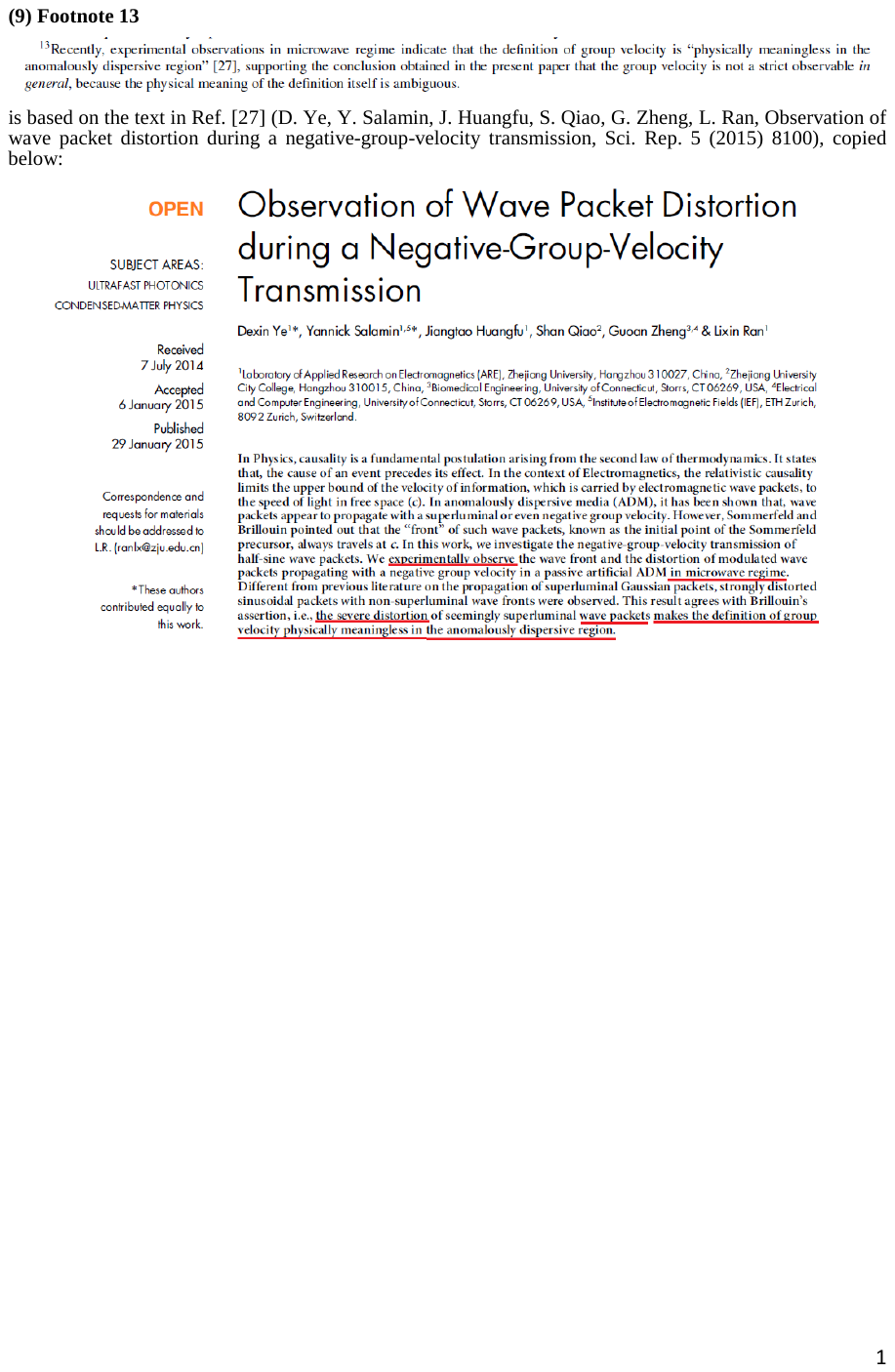}
%%\caption{}
\label{figM12}
\end{figure}

\end{document}